%% file: main.tex
\documentclass[preprint]{elsarticle} 

\makeatletter
	\def\ps@pprintTitle{%
 	\let\@oddhead\@empty
	\let\@evenhead\@empty
	\def\@oddfoot{\centerline{\thepage}}%
	\let\@evenfoot\@oddfoot}
\makeatother

\usepackage{etoolbox}
\patchcmd{\MaketitleBox}{\footnotesize\itshape\elsaddress\par\vskip36pt}{\footnotesize\itshape\elsaddress\par\parbox[b][36pt]{\linewidth}{\vfill\hfill\textnormal{\today}\hfill\null\vfill}}{}{}%
\patchcmd{\pprintMaketitle}{\footnotesize\itshape\elsaddress\par\vskip36pt}{\footnotesize\itshape\elsaddress\par\parbox[b][36pt]{\linewidth}{\vfill\hfill\textnormal{\today}\hfill\null\vfill}}{}{}%
\usepackage{multirow}
\usepackage{tikz}
\usetikzlibrary{shapes.geometric, arrows}

\tikzstyle{startstop} = [rectangle, 
minimum width=6cm, 
minimum height=1cm,
text centered, 
draw=black, 
text width=6cm, 
fill=white!20]

\tikzstyle{arrow} = [thick,->,>=stealth]
            
\input{ListPackages}
\input{mymacros.tex}

\begin{document}

\begin{frontmatter}
\title{Bayesian Inference and Global Sensitivity Analysis for Ambient Solar Wind Prediction}

\author[1]{Opal Issan{\corref{cor1}}}\ead{oissan@ucsd.edu}
\cortext[cor1]{Corresponding author}
\author[2]{Pete Riley}
\author{Enrico Camporeale\texorpdfstring{$^{\mathrm{c}, \mathrm{d}}$}}
\author[1]{Boris Kramer}

\address[1]{Department of Mechanical and Aerospace Engineering, University of California San Diego, CA, United States}
\address[2]{Predictive Science Inc., San Diego, CA, USA}
\address[3]{NOAA Space Weather Prediction Center, Boulder, CO, USA}
\address[4]{CIRES, University of Colorado, Boulder, CO, USA}

\begin{abstract}
The ambient solar wind plays a significant role in propagating interplanetary coronal mass ejections and is an important driver of space weather geomagnetic storms. 
A computationally efficient and widely used method to predict the ambient solar wind radial velocity near Earth involves coupling three models: Potential Field Source Surface, Wang-Sheeley-Arge (WSA), and Heliospheric Upwind eXtrapolation.
However, the model chain has eleven uncertain parameters that are mainly non-physical due to empirical relations and simplified physics assumptions. 
We, therefore, propose a comprehensive uncertainty quantification (UQ) framework that is able to successfully quantify and reduce parametric uncertainties in the model chain. The UQ framework utilizes variance-based global sensitivity analysis followed by Bayesian inference via Markov chain Monte Carlo to learn the posterior densities of the most influential parameters. 
The sensitivity analysis results indicate that the five most influential parameters are all WSA parameters. Additionally, we show that the posterior densities of such influential parameters vary greatly from one Carrington rotation to the next. The influential parameters are trying to overcompensate for the missing physics in the model chain, highlighting the need to enhance the robustness of the model chain to the choice of WSA parameters. The ensemble predictions generated from the learned posterior densities significantly reduce the uncertainty in solar wind velocity predictions near Earth.
\end{abstract}	
\end{frontmatter}

\section{Introduction}
The ambient (or background) solar wind is the long-lived large-scale plasma that emanates from the Sun and travels into interplanetary space, which excludes interplanetary coronal mass ejections and other transient events.
It is crucial to accurately predict the ambient solar wind since interplanetary coronal mass ejections, the primary source of extreme space weather events, are modeled as perturbations to the ambient solar wind~\cite{odstrcil_1999_icme}. Additionally, corotating interaction regions between fast and slow ambient solar wind streams are drivers of moderate space weather events~\cite{riley_2012_cir}. In fact, corotating interaction regions have been found to contribute to 70\% of geomagnetic activity at Earth during solar minimum and about 30\% during solar maximum~\cite{richardson_2000_sources}. Thus, generating reliable predictions of the ambient solar wind is essential for improving space weather prediction capabilities and for accurately assessing the risk of space weather events.

State-of-the-art ambient solar wind models couple two regions: the corona and heliosphere. The coronal domain spans from the surface of the Sun (1$R_{\mathrm{S}}$, i.e., one solar radius) up to the coronal outer boundary, which is typically set to a distance between $2.5 R_{\mathrm{S}}$ to $30 R_{\mathrm{S}}$ depending on the model that is used. The solution at the coronal outer boundary is extrapolated into the heliospheric domain up to Earth's orbit and beyond. High-fidelity simulations of the ambient solar wind are constructed via time-dependent magnetohydrodynamic (MHD) models, such as the Magnetohydrodynamics Algorithm outside a Sphere~\cite{linker_1999_mas, mikic_2018_eclipse, riley_2019_psp} and the Space Weather Modeling Framework~\cite{toth_2005_swmf, van_der_holst_2014_awsom}. Such MHD models simulate the ambient solar wind by relaxing the coronal and heliospheric simulations to a steady-state solution, requiring high computational costs. In an effort to reduce simulation time (especially in operational settings), the space weather community commonly uses lower-fidelity models based on reduced physics and empirical relations. A well-established and widely-used chain of lower-fidelity models couples the Potential Field Source Surface (PFSS)~\cite{pfss_altschuler_1969, pfss_schatten_1969}, Wang-Sheeley-Arge (WSA)~\cite{arge_2004}, and Heliospheric Upwind eXtrapolation (HUX)~\cite{riley_HUXP1_2011, riley_HUXP2_2021, issan_HUXP3_2022} models. 

There has been a continuous effort in the space weather community to improve ambient solar wind models by comparing the model predictions to \textit{in-situ} spacecraft observations~\cite{reiss_2022_validation_ambient}. Such \textit{in-situ} observations can also be leveraged to reduce prediction uncertainties stemming from initial conditions, boundary conditions, fitting parameters, numerical errors, measurement noise, etc. Here, we study parametric uncertainty in the PFSS$\to$WSA$\to$HUX model chain. We rigorously examine the uncertainty and sensitivity of eleven parameters, such as the source surface height and WSA numerical parameters, and their impact on the solar wind radial velocity predictions near Earth. Quantifying and reducing the parametric uncertainties in the ambient solar wind models is critical for making informed decisions in operational settings. 

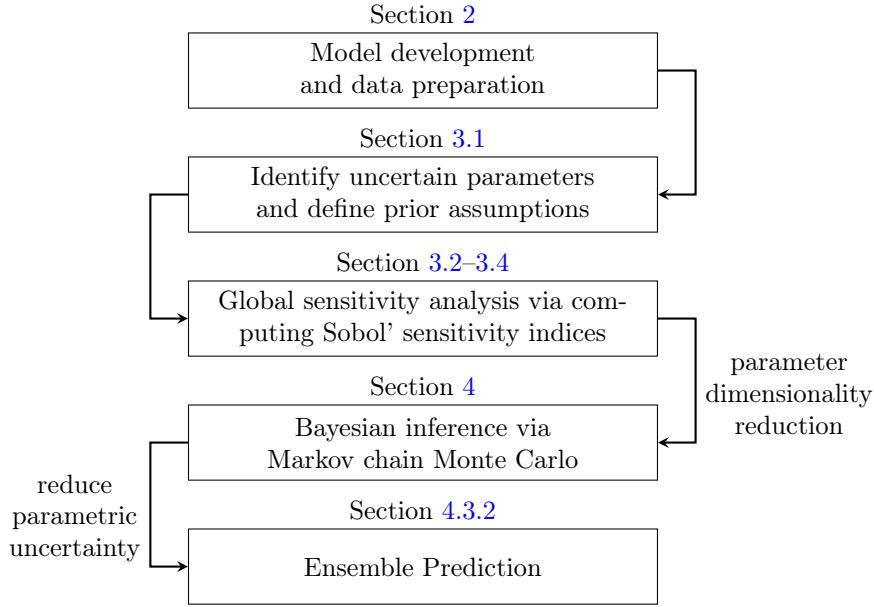
\begin{figure}
    \centering
    \begin{tikzpicture}[node distance=1.15cm]
    \node (model) [startstop, label=above:Section~\ref{sec:models}] {Model development and data preparation};
    \node (prior) [startstop, below of=model, yshift=-0.5cm, label=above:Section~\ref{sec:uncertain-parameters}] {Identify uncertain parameters and define prior assumptions};
    \node (gsa) [startstop, below of=prior, yshift=-0.5cm, label=above:Section~\ref{sec:sobol}--\ref{sec:gsa-results}] {Global sensitivity analysis via computing Sobol' sensitivity indices};
    \node (mcmc) [startstop, below of=gsa, yshift=-0.5cm, label=above:Section~\ref{sec:bayes-and-mcmc}] {Bayesian inference via Markov chain Monte Carlo};
    \node (ensemble) [startstop, below of=mcmc, yshift=-0.5cm, label=above:Section~\ref{sec:Ensemble-results}] {Ensemble Prediction};
    
    \draw[arrow] (model.east) -| ++(.5,-1) node[anchor=west, align=center] {} |-  (prior.east) ;
    \draw[arrow] (prior.west) -| ++(-.5, -1) node[anchor=east, align=center] {} |-  (gsa.west) ;
    \draw[arrow] (gsa.east) -| ++(.5,-1) node[anchor=west, align=center] {parameter\\dimensionality\\reduction} |-  (mcmc.east) ;
    \draw[arrow] (mcmc.west) -| ++(-.5, -1) node[anchor=east, align=center] {reduce\\parametric\\uncertainty} |-  (ensemble.west) ;
    \end{tikzpicture}
    \caption{A comprehensive UQ framework that allows for ensemble predictions with reduced uncertainties. The framework proceeds in the following steps: First, we identify the problem space and parameters that have uncertainties. Second, through global sensitivity analysis, we find out which of these parameters significantly impacts the variance of the quantity of interest. Third, we infer the full distribution of the most important parameters through Bayesian inference. Fourth, we make ensemble predictions with these newfound parameter distributions. }
    \label{fig:uq-flowchart}
\end{figure}

As a core contribution of this work, we present a comprehensive uncertainty quantification (UQ) framework to advance the use of rigorous UQ techniques in space weather. The proposed UQ framework is described in the following steps and illustrated in Figure~\ref{fig:uq-flowchart}. First, we perform variance-based global sensitivity analysis to identify which parameters influence the solar wind predictions near Earth the most. Subsequently, parameters that hardly contribute to the prediction variability are set to their fixed deterministic values, which facilitates \textit{a posteriori} parameter dimensionality reduction. Then, we apply Bayesian inference to uncover the posterior density of the influential parameters, which is the conditional probability of the influential parameters given observational data. Lastly, we sample from the learned posterior densities and generate an ensemble of the ambient solar wind predictions near Earth, which demonstrates that the UQ framework reduces the parametric uncertainty in the predicted solar wind velocity.

Sensitivity analysis quantifies the contribution of parametric uncertainty on the variability of a quantity of interest (QoI). 
Common methods can be classified into two groups: local and global. Local sensitivity analysis methods vary the parameters about a deterministic value by computing local partial derivatives, whereas global sensitivity analysis methods account for variance effects in the entire parameter space~\cite{saltelli_2008_gsa_book}.
We use variance-based global sensitivity analysis by estimating Sobol' sensitivity indices~\cite{sobol_2001_theory}. A sensitivity analysis study by~\cite{jivani_2022_gsa} estimated the Sobol' sensitivity indices associated with uncertain parameters in the MHD Alfv\'en Wave Solar atmosphere Model~\cite{van_der_holst_2014_awsom}. A recent study by~\cite{reiss_2020_parametric} used Morris screening~\cite{morris_1991_ee}, a hybrid local/global sensitivity analysis method, to identify the most influential parameters in the WSA model. The Morris screening method averages local derivative approximations to provide global sensitivity measures~\cite[\S 15.2]{ralph_smith_uq_2013}. It is typically used when variance-based methods are prohibitively expensive since it can only rank the parameters based on their importance but, unlike variance-based methods, does not quantify the relative contributions of each parameter to the QoI variance. Our study differs from~\cite{reiss_2020_parametric} since we consider parametric uncertainty stemming not only from WSA but also from PFSS and HUX. Additionally, our study differs from~\cite{reiss_2020_parametric} since we compute full global sensitivity information.

Parameter estimation techniques are typically divided into two approaches: frequentist and Bayesian. The frequentist approach seeks to find a single `optimal' value of each parameter by solving an optimization problem. For example, maximum likelihood estimation~\cite{milton_2003_statistics}, such that the optimal values of the parameters minimize the difference between the model prediction and observational data. 
\cite{riley_2015_parametric}, \cite{reiss_2020_parametric}, and \cite{kumar_2022_parametric} took a frequentist approach to find the optimal parameters in the WSA model. This paper presents a Bayesian approach to learning the uncertain parameters in the PFSS$\to$WSA$\to$HUX model chain.
The Bayesian approach views the parameters as random variables and seeks to learn their posterior density.
In the Bayesian setting, the solution to the UQ inverse problem is represented by a probability density function of the parameters. In stark contrast, the frequentist approach yields a point estimate. 
Thus, the Bayesian approach provides a complete picture of the uncertainty associated with the model parameters. Subsequently, relevant point estimates, such as the maximum \textit{a posteriori} (MAP), variance, and mean, can all be computed from the posterior density.
Here, we use Markov chain Monte Carlo (MCMC)~\cite{metropolis_1953_mcmc, hastings_1970_mcmc} to learn the posterior densities of the most influential parameters. In particular, we employ the MCMC affine invariant ensemble sampler~\cite{goodman_2010_emcee, mackey_python_2013_emcee}, which is robust to different scales in the parameters. 
 
The main questions we seek to answer via the proposed UQ framework are: (1) How does parametric uncertainty in the PFSS$\to$WSA$\to$HUX model chain impact the uncertainty in the solar wind velocity predictions near Earth? Can we reduce such uncertainties using Bayesian inference methods? (2) What are the most influential parameters in the model chain? (3) How do the posterior densities of the influential parameters change over time? Is there a clear trend in the posterior evolution? (4) Is the model chain robust to the choice of its parameters? Is it reliable enough to be used for real-time operational forecasting? 

This paper is organized as follows. Section~\ref{sec:models} describes the models and observational data used in this work. Section~\ref{sec:gsa} discusses variance-based global sensitivity analysis, an algorithm to compute Sobol' sensitivity indices via Monte Carlo integration, and numerical results. In Section~\ref{sec:bayes-and-mcmc} we discuss Bayesian inference via MCMC algorithms and numerical results. Section~\ref{sec:conclusion} then offers conclusions and an outlook to future work.

\section{Ambient Solar Wind Model Chain and Observational Data} \label{sec:models}
We consider the coupling of three well-established models: (1) Potential Field Source Surface (PFSS), (2) Wang-Sheeley-Arge (WSA), and (3) Heliospheric Upwind eXtrapolation (HUX), to predict the ambient solar wind radial velocity near Earth. \cite{reiss_2019_parametric, reiss_2020_parametric} and \cite{bailey_2021_model_chain} use a similar chain of models with the addition of the Schatten Current Sheet (SCS) model, developed by~\cite{pfss_schatten_1969}, resulting in the following model chain: PFSS$\to$SCS$\to$WSA$\to$HUX. The SCS model is added to correct the PFSS radial magnetic field latitudinal variations to match Ulysses' observations~\cite{wang_1995_ulysses}. However, a recent study by \cite{kumar_2022_parametric} showed that adding SCS to the chain of models did not necessarily improve the accuracy of the solar wind radial velocity predictions at L1 (during 2006-2011). We thus analyze the PFSS$\to$WSA$\to$HUX model chain.

\begin{figure}
    \centering
    \includegraphics[width=\textwidth]{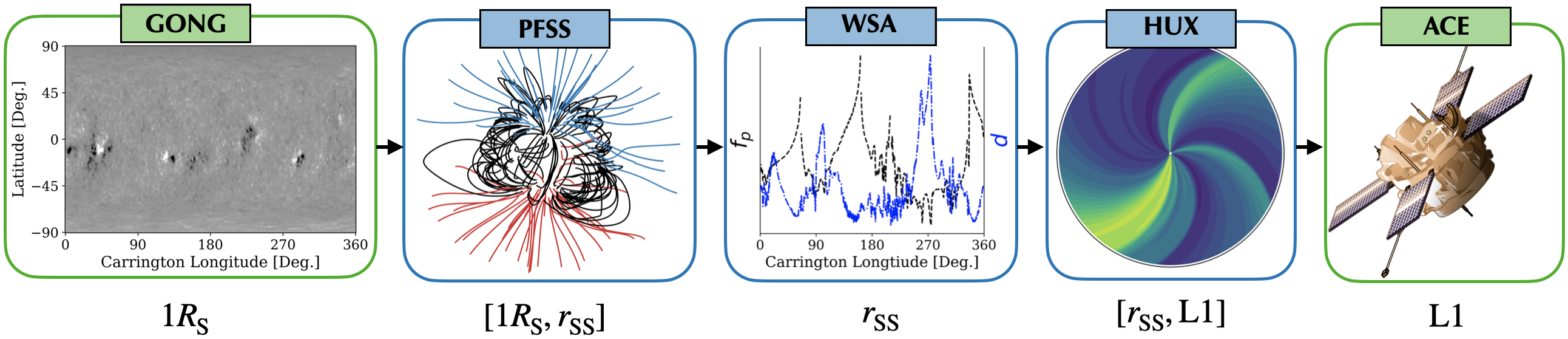}
    \caption{A flowchart of the models (blue panels: PFSS, WSA, and HUX) and observational data (green panels: GONG and ACE) utilized in this study. The GONG and ACE images are adapted from NSO and NASA, respectively. }
    \label{fig:model-chain-graphical-abstract}
\end{figure}

The PFSS model is used to predict the magnetic field in the coronal domain. The PFSS magnetic field solution is then used as an input to the WSA relation, which computes the solar wind radial velocity at the outer boundary of the coronal domain. The WSA results are then set as the initial condition for the HUX model, which extrapolates the solar wind radial velocity into the heliospheric domain. Finally, the model chain solar wind radial velocity predictions are compared with ACE spacecraft \textit{in-situ} observations. We use synoptic magnetograms from Global Oscillation Network Group (GONG) as the inner boundary condition for the PFSS model. A flowchart of the models and data used in this study is shown in Figure~\ref{fig:model-chain-graphical-abstract}. The subsequent sections explain the different components of the model chain and observational data in further detail. 

\begin{figure}
    \centering
    \includegraphics[width=\textwidth]{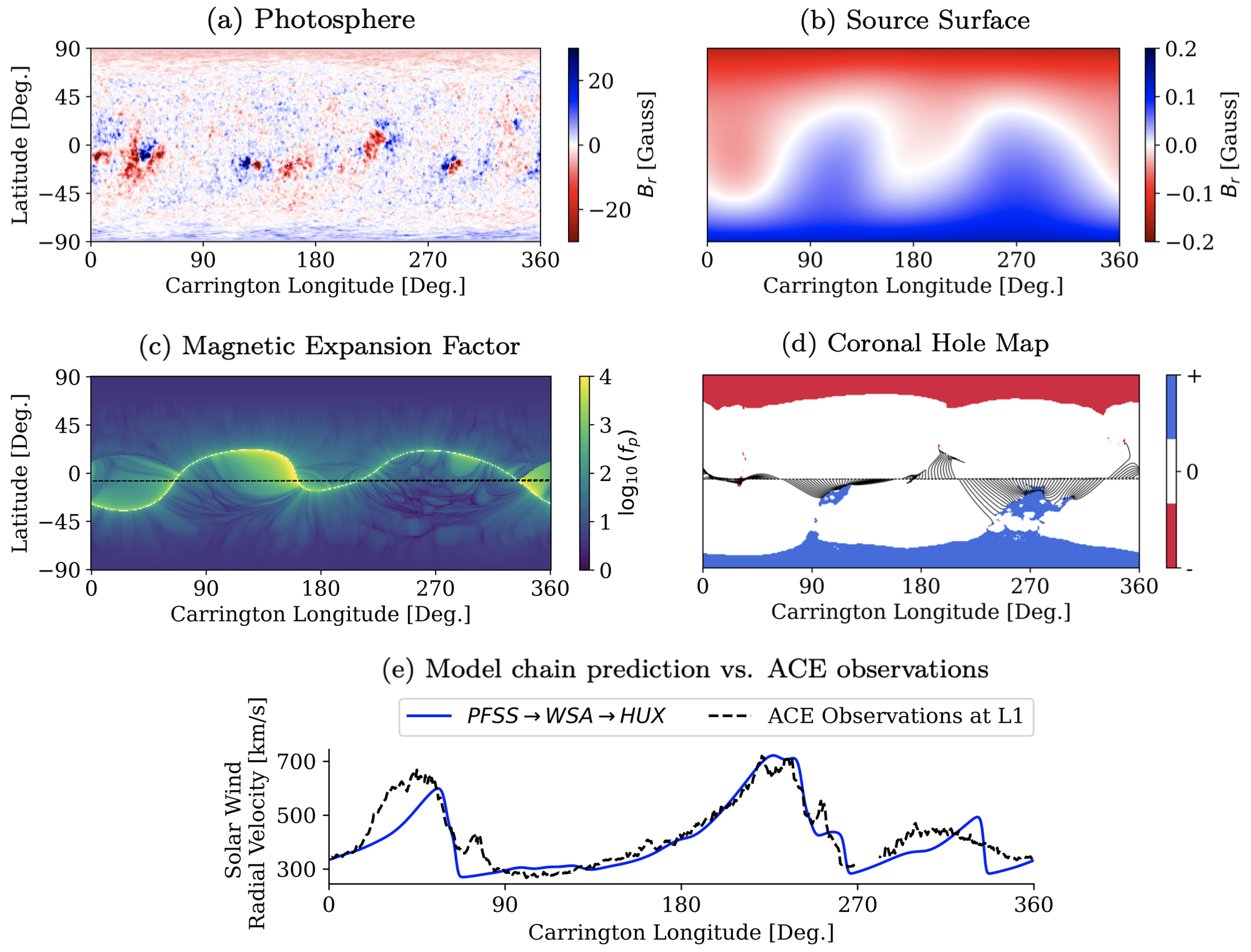}
    \caption{An illustration of different components in PFSS$\to$WSA$\to$HUX model chain during CR 2053. The PFSS radial magnetic field at the lower and upper boundaries is shown in (a) and (b). The lower boundary condition is obtained from GONG synoptic maps and the extrapolated upper boundary is shown at the source surface, which is set to $r_{\mathrm{SS}} = 2.5 R_{\mathrm{S}}$ in this example. The WSA inputs, i.e. the magnetic field expansion factor and coronal hole map, are shown in (c) and (d). In subfigure~(d), the red (blue) coronal hole areas show the negative inward (positive outward) fields and the black dashed lines show the magnetic field lines traced from ACE's projected trajectory at the source surface  back to the photosphere. Lastly, (e) shows a comparison of the model chain solar wind radial velocity predictions at L1 with ACE \textit{in-situ} observations. In this example, we set the WSA parameters to $v_{0} = 250 \frac{\mathrm{km}}{\mathrm{s}}, v_{1} = 945 \frac{\mathrm{km}}{\mathrm{s}}, \alpha = 0.16, \beta = 1, \gamma = 0.6, w = 0.02 \mathrm{rad}, \delta = 1.75, \psi = 3$ and the HUX parameters to $\alpha_{\mathrm{acc}} = 0.15, r_{h} = 50 R_{\mathrm{S}}$.}
    \label{fig:model-components-CR2053}
\end{figure}


\subsection{Potential Field Source Surface (PFSS) Model} \label{sec:pfss}
The PFSS model proposed by~\cite{pfss_altschuler_1969} and~\cite{pfss_schatten_1969} solves for the coronal magnetic field $\mathbf{B}(r, \theta, \phi)=[B_{r}(r, \theta, \phi), B_{\theta}(r, \theta, \phi), B_{\phi}(r, \theta, \phi)]$ from the photosphere (the visible surface of the Sun) to the outer radius called the \textit{source surface}. The PFSS model assumes that beyond the source surface, the magnetic field is purely radial, i.e. open magnetic field lines are carried into interplanetary space by the solar wind. 
Additionally, the PFSS model neglects the coronal electric current density since, above the photosphere, there is a large decrease in particle density and a smaller decrease in magnetic field strength~\cite{kruse_2020_pfss}; it also assumes that the corona is electrostatic since during solar minimum, the corona evolves slowly, and features can last for several Carrington rotations (CRs). It is important to mention that some of the above assumptions hold less during specific time periods. For example, during solar maximum, the photospheric field changes more rapidly, challenging the electrostatic assumption. Additionally, \cite{riley_2006_pfss_vs_mhd} found that the concept of spherical source surface is more reasonable during solar maximum than during solar minimum. 
These assumptions (coupled with Amp\`ere's law) lead to 
\begin{equation*}
        \nabla \times \mathbf{B} = 0,
\end{equation*}
so that the magnetic field can be described by its potential $\mathbf{B} = -\nabla \Psi$. By combining the potential description with Gauss's law ($\nabla \cdot \mathbf{B} = 0$), we get Laplace's equation
\begin{equation*}
    \nabla^2 \Psi = 0, 
\end{equation*}
subject to the following boundary conditions
\begin{equation}\label{boundary_conditions_pfss}
    \begin{split}
    \frac{\partial \Psi}{\partial r}(r=1 R_{\mathrm{S}}, \theta, \phi) &= g(\theta, \phi),\\
    \frac{\partial \Psi}{\partial \theta} (r = r_{\mathrm{SS}}, \theta, \phi) &= \frac{\partial \Psi}{\partial \phi} (r=r_{\mathrm{SS}}, \theta, \phi) = 0,\\
    \Psi(r, \theta, \phi=0) &= \Psi(r, \theta, \phi=2\pi),
    \end{split}
\end{equation}
where $\theta \in [0, \pi]$ is Carrington colatitude, $\phi \in [0, 2\pi]$ is Carrington longitude, $R_{\mathrm{S}}$ denotes solar radii unit of distance which is 695,700km, and approximately 1/215th of an astronomical unit (AU), $r \in [1R_{\mathrm{S}}, r_{\mathrm{SS}}]$ is the radial distance from the center of the Sun, $r_{\mathrm{SS}}$ is the source surface height, and $g(\theta, \phi)$ is a given photospheric synoptic map. 
The PFSS model is typically solved via spherical harmonic expansion or numerical discretization methods \cite{caplan_2021_finite_difference_pfss, liu_2022_discontinious_galerkin, stansby_2020}.
We employ the \texttt{pfsspy} Python package (version 1.1.2), developed by~\cite{stansby_2020}, for solving PFSS via finite-difference discretization and for tracing magnetic field lines. The finite-difference discretization is on a rectilinear grid equally spaced in $\sin$(colatitude), longitude, and $\ln$(radius) coordinates, see~\cite{stansby_2020} for more details on the solver. In this study, all simulations are performed on a $180 \times 360 \times 100$ grid resolution in $\sin$(colatitude), longitude, and $\ln$(radius), respectively, i.e. we solve for $6.48 \times 10^{6}$ states.
As an illustrative example, Figure~\ref{fig:model-components-CR2053}(a) shows the radial magnetic field at the photosphere (inner boundary) for CR 2053 obtained by the GONG synoptic maps (see Section~\ref{sec:gong}) and Figure 1(b) shows the radial magnetic field results at the source surface (outer boundary), which is set to $r_{\mathrm{SS}} = 2.5 R_{S}$ for this example. 

\subsection{Wang-Sheeley-Arge (WSA) Model} \label{sec:wsa}
The WSA model developed by~\cite{arge_2004} is a semi-empirical model of the ambient solar wind velocity in the inner-heliosphere, which fuses the Wang-Sheeley (WS) model developed by~\cite{wang_sheeley_1990_ws} with the distance to the coronal hole boundary (DCHB) model developed by~\cite{riley_2001_dchb}. 
The WSA model (coupled with the MHD Enlil model) is used in operational forecasting at the National Oceanic and Atmospheric Administration (NOAA) Space Weather Prediction Center~\cite{parsons_operational_wsa_enlil_2011}.
The WSA model is given by 
\begin{equation*} 
    v_{\mathrm{wsa}}(f_{p}, d; v_{0}, v_{1}, \alpha, \beta, \gamma, w, \delta, \psi) = v_{0} + \frac{v_{1} - v_{0}}{(1+f_{p})^{\alpha}} \left(\beta - \gamma \exp \left(-\left(\frac{d}{w}\right)^{\delta} \right)\right)^{\psi},
\end{equation*}
where $v_{0}$ and $v_{1}$ correspond to the minimum and maximum solar wind velocities, $d$ is the minimum angular distance that an open field footpoint lies from a coronal hole boundary, $f_{p}$ is the magnetic field expansion factor, and $\alpha, \beta, \gamma, \delta, w, \psi$ are additional tunable parameters.

The magnetic expansion factor $f_{p}$ is derived from the coronal magnetic field by tracing down field lines from the source surface to the photosphere, namely
\begin{equation} \label{expansion_factor}
    f_{p} = \left(\frac{1R_{\mathrm{S}}}{r_{\mathrm{SS}}}\right)^2\left|\frac{B_{r}(1R_{\mathrm{S}}, \theta_{\mathrm{p}}, \phi_{\mathrm{p}})}{B_{r}(r_{\mathrm{SS}}, \theta_{\mathrm{SS}}, \phi_{\mathrm{SS}})}\right|,
\end{equation}
where $B_{r}(r, \theta, \phi)$ is the radial magnetic field component, and the subscripts $\mathrm{p}$ and $\mathrm{SS}$ refer to the field line trace at the photosphere and solar surface, respectively.
The distance to the coronal hole boundary $d$ is also derived from the coronal magnetic field solution via a two-step approach.
First, the coronal hole regions are identified by tracing field lines from the photosphere to the source surface and detecting the footpoint of all open magnetic field lines, i.e. coronal hole regions. 
Second, the great-circle angular distance $d$ is computed between the footpoints of the open magnetic field lines to the nearest coronal hole boundary.
To illustrate these concepts, Figure~\ref{fig:model-components-CR2053}(c) presents the magnetic expansion factor for CR~2053, and the black dashed line shows ACE's spacecraft projected trajectory. Similarly, Figure~\ref{fig:model-components-CR2053}(d) shows the coronal hole map for CR 2053 with ACE's trajectory field line traces, which mainly trace down to low-latitude coronal holes. 

\subsection{Heliospheric Upwind eXtrapolation (HUX) Model} \label{sec:hux}
The two-dimensional HUX model developed by~\cite{riley_HUXP1_2011} extrapolates the coronal solar wind radial velocity into the heliospheric domain. The HUX model is based on simplified physical assumptions of the fluid momentum equation, which reduces to the following nonlinear scalar homogeneous time-stationary equation
\begin{equation} \label{hux-underlying-equation-PDE}
-\Omega_{\mathrm{rot}}(\theta = \hat{\theta}) \frac{\partial v(r, \phi)}{\partial \phi} + v(r, \phi)\frac{ \partial v(r, \phi)}{\partial r}=0,
\end{equation}
where the independent variables are the radial distance from the Sun $r$ and Carrington longitude $\phi$, and the dependent variable is the solar wind radial velocity $v(r, \phi)$. The angular frequency of the Sun's rotation is evaluated at a constant Carrington colatitude $\hat{\theta}$~\cite{riley_HUXP2_2021}, which is estimated by 
$\Omega_{\mathrm{rot}}(\theta) = \frac{2 \pi}{25.38} - \frac{2.77 \pi}{180} \cos \left(\frac{\pi}{2} - \theta \right)^2$.
The problem is subject to the boundary condition $v(r_{\mathrm{SS}}, \phi) = v_{r_{\mathrm{SS}}} (\phi)$ and is defined on the longitudinal periodic domain $0 \leq \phi \leq 2 \pi$ and $r \geq  r_{\mathrm{SS}}$. \cite{riley_HUXP1_2011} suggest adding an acceleration boost to the boundary condition (before propagation) to account for the residual acceleration present in the inner heliosphere, i.e. 
\begin{equation} \label{hux-acceleration-boost-term}
v_{\mathrm{acc}}(r_{\mathrm{SS}}, v_{r_{\mathrm{SS}}} (\phi); \alpha_{\mathrm{acc}}, r_{h}) = \alpha_{\mathrm{acc}}( 1 - e^{-r_{\mathrm{SS}}/r_{h}})v_{r_{\mathrm{SS}}} (\phi),
\end{equation}
where $v_{r_{\mathrm{SS}}} (\phi)$ is the radial velocity at the source surface (obtained from WSA relation), $\alpha_{\mathrm{acc}}$ is the acceleration factor, and $r_{h}$ is the radial location at which the acceleration ends. We discretize Eq.~\eqref{hux-underlying-equation-PDE} via finite-differencing on a uniform mesh with $600 \times 300$ resolution in $\phi \in [0, 2\pi]$ and $r \in [r_{\mathrm{SS}}, r_{\mathrm{max}}]$, respectively. We set $r_{\mathrm{max}}$ to be ACE's maximum radial distance from the Sun for the considered CR. We solve the equation using the first-order upwind scheme, see~\cite{issan_HUXP3_2022} for more details about the numerical scheme and stability requirements. Figure~\ref{fig:model-components-CR2053}(e) shows the coupling of PFSS, WSA, and HUX solar wind speed predictions in comparison to ACE's \textit{in-situ} observations for CR 2053, see Section~\ref{sec:ace} for more details about ACE. We set $\hat{\theta}$ to be ACE's mean latitude over a CR and obtain the inner boundary velocity profile by computing the magnetic field expansion and distance to the coronal hole at ACE's projected trajectory (which are inputs in the WSA model). The solar wind velocity at ACE's trajectory is obtained by linearly interpolating the two-dimensional HUX solution along ACE's trajectory.

\subsection{Observational Data} \label{sec:data}
\subsubsection{Global Oscillations Network Group (GONG) Synoptic Magnetograms} \label{sec:gong}
Deployed in 1995, the GONG synoptic magnetograms are produced every hour at GONG's six ground-based sites with identical telescopes. The six sites in California, Hawaii, Australia, India, Spain, and Chile, are distributed worldwide so that the Sun is visible  at nearly all times. 
The line-of-sight full-disk GONG magnetograms are provided every minute by its main instrument known as Fourier Tachometer~\cite{hill_2018_gong, harvey_1996_gong}.
The magnetic field strength (measured in Gauss) is determined spectroscopically using the Zeeman effect. In the presence of a magnetic field, gas spectral lines split into two or more components, and the frequency of the spectral lines depends on the strength of the magnetic field~\cite{moldwin_2008_space_weather_book}.

In this study, we use the GONG full CR synoptic magnetograms as the photospheric radial magnetic field $g(\theta, \phi)$ boundary condition for the PFSS model, see Eq.~\eqref{boundary_conditions_pfss}, which are publicly available at National Solar Observatory's website\footnote{\url{https://gong.nso.edu/data/magmap/crmap.html}}. The synoptic maps are calibrated from roughly 8,000-10,000 input full-disk 10-min average magnetograms~\cite{hill_2018_gong} and are provided at $180 \times 360$ resolution in Carrington $\sin$(colatitude) and Carrington longitude, respectively. These synoptic maps are obtained over a full CR and reasonably approximate the solar conditions at quiet times of the cycle when the solar evolution is slow. Figure~\ref{fig:model-components-CR2053}(a) above shows the GONG synoptic map for CR 2053. 

\subsubsection{Advanced Composition Explorer (ACE) \textit{in-situ} Solar Wind Measurements} \label{sec:ace}
The NASA ACE satellite launched in 1997 is in Lissajous orbit around L1 (one of Earth-Sun gravitational equilibrium points), located about $1.5 \times 10^{6}$~km forward of Earth. The location of ACE gives about 1-hour advance warning of the arrival of space weather events on Earth. The ACE instruments measure the solar wind, interplanetary magnetic field, and high-energy particles. This study uses the solar wind radial velocity \textit{in-situ} measurements provided by ACE's Solar Wind Electron Proton Alpha Monitor (SWEPAM) instrument~\cite{mccomas_1998_ace}. To download the radial velocity data and ACE's trajectory at a 1-hour cadence, we used \texttt{HelioPy}, a community-developed Python package~\cite{david_stansby_2020_heliopy}, for retrieving space physics datasets from NASA’s Space Physics Data Facility website\footnote{\url{https://cdaweb.gsfc.nasa.gov/index.html/}}.

\subsection{Model Chain Simulations} 
The PFSS$\to$WSA$\to$HUX model chain simulations are run on the \emph{Alfv\'en} server at the University of Colorado SWx-TREC (Space Weather Technology, Research, and Education Center), which is equipped with 2x AMD EPYC 74F3 24-Core processors (3.2 GHz) and a total 2 Tb of RAM. The model chain takes about 16 seconds to simulate on one CPU. We profiled the model chain computations and found that 98\% of the total time is spent solving the PFSS model and computing the distance to the coronal hole and magnetic expansion, 1.8\% is spent on solving the HUX model and less than a percent is spent on evaluating the WSA model.The sensitivity analysis results required $3 \times 1.3 \times 10^{5} = 3.9 \times 10^{5}$ simulations and the MCMC results required $10 \times 250 \times 2.6 \times 10^{4} = 6.5 \times 10^{7}$ model simulations, i.e. a total of approximately $10{,}400$ CPU hours.

\section{Global Sensitivity Analysis} \label{sec:gsa}
Variance-based global sensitivity analysis aims to identify the parameters that contribute the most to a given QoI variability, which can be done quantitatively via computing Sobol' sensitivity indices~\cite{sobol_2001_theory}. 
Parameters with high sensitivity indices are classified as influential, whereas parameters with low sensitivity indices are classified as non-influential. 
Computing Sobol' sensitivity indices facilitate \textit{a posteriori} parameter dimensionality reduction in subsequent inverse UQ tasks (such as Bayesian inference). 
This is established by setting non-influential parameters to their deterministic values and only considering parametric uncertainty stemming from influential parameters. 
Parametric dimensionality reduction is often necessary for computationally demanding models and unbiased inverse UQ methods. 

\subsection{Uncertain Parameters} \label{sec:uncertain-parameters}
The PFSS$\to$WSA$\to$HUX model chain has many parameters that are uncertain, inducing uncertainty in the solar wind velocity forecasts near Earth. All uncertain parameters in the model chain are mainly non-physical. We identified a total of eleven uncertain continuous parameters: one parameter in PFSS (source surface height), eight parameters in WSA (numerical parameters), and two parameters in HUX (acceleration parameters). Table~\ref{tab:uncertain-parameters} lists the uncertain input parameters and their corresponding prior densities. We set all prior densities to be uniform with reasonable ranges determined in previous parametric studies by~\cite[\S 2]{lee_2011_source_surface}, \cite[\S 2.5]{arden_2014_source_surface}, \cite[Eq. 9]{meadors_2020_particle_filter}, \cite[Table 1]{kumar_2022_parametric}, and \cite[Table 1]{riley_2015_parametric}.

The source surface radial height $r_{\mathrm{SS}}$ in the PFSS model has intrinsic uncertainties since, in reality, it is non-spherical and is a function of space and time. 
Lowering the source surface results in more coronal holes, open flux, and strong curvature in the heliospheric current sheet, whereas raising the source surface height results in the opposite effect. 
\cite{riley_2006_pfss_vs_mhd} suggested avoiding the strict constraint of a spherical source surface by a detailed comparison of PFSS to MHD models, and \cite{kruse_2020_pfss} altered the PFSS model to employ an oblate or prolate elliptical source surface. 
\cite{arden_2014_source_surface} show that the source surface has a ``breathing" effect of which the canonical $2.5R_{S}$ source surface, originally suggested by~\cite{pfss_altschuler_1969}, matches measured interplanetary magnetic field (IMF) open flux near Earth during solar maximum, yet extends up to $4R_{S}$ during solar minimum of solar cycle 23 and the start of cycle 24. 
A similar study by~\cite{lee_2011_source_surface} found that setting the source surface to $1.8R_{S}$ matched best the IMF strength during the minimum of solar cycle 23. 
The optimal source surface heights determined in~\cite{arden_2014_source_surface} and ~\cite{lee_2011_source_surface} do not agree and further emphasize the need for additional numerical investigation.
Additionally, \cite[Figure~14]{lee_2011_source_surface} and \cite[Figure~3]{nikolic_2019_source_surface} 
compared the PFSS coronal holes to observed extreme ultraviolet synoptic images, their results suggest $1.5-1.8R_{S}$ for the source surface during CR 2060.
Similar to our study, \cite{meadors_2020_particle_filter} also considers the source surface as an uncertain input parameter and learns its density via particle filtering and WIND spacecraft observations.
Based on \cite[\S 2]{lee_2011_source_surface}, \cite[\S 2.5]{arden_2014_source_surface} and \cite[Eq. 9]{meadors_2020_particle_filter}, we allow the source surface to vary from $1.5R_{\mathrm{S}}$ to $4R_{\mathrm{S}}$.

The eight numerical parameters of the WSA model, $v_{0}, v_{1}, \alpha, \beta, \gamma, \delta, w, \psi$, similar to the source surface, cannot be directly measured and are usually adjusted for different observatories, e.g. Wilcox solar observatories and GONG~\cite{riley_2015_parametric}. Additionally, \cite{riley_2015_parametric} and \cite{kumar_2022_parametric} showed that the optimal parameter can vary greatly from one CR to the next. It is, therefore, important to understand the uncertainties in the WSA parameters and their impact on predicted solar wind speed near Earth. We set the eight parameter ranges based on previous parametric studies by~\cite[Table 1]{riley_2015_parametric} and~\cite[Table 1]{kumar_2022_parametric}.

The HUX model has two free parameters $\alpha_{\mathrm{acc}}$ and $r_{h}$ in the acceleration boost term, see Eq.~\eqref{hux-acceleration-boost-term}. \cite{riley_HUXP1_2011} suggest setting $\alpha_{\mathrm{acc}} = 0.15$ and $r_{h} = 50 R_{\mathrm{S}}$. A recent study by~\cite{riley_HUXP2_2021} compared HUX to three-dimensional MHD velocity predictions and found the optimal $\alpha_{\mathrm{acc}}$ and $r_{h}$ via nonlinear least-squares optimization for a few CRs spanning from CR 2029 to CR 2231. They found that the average optimal $\alpha_{\mathrm{acc}}$ and $r_{h}$ are $0.16$ and $52.6 R_{S}$, respectively. \cite{riley_HUXP2_2021} took a frequentist approach to find the optimal HUX parameters. In this study, we formulate the inference problem using the Bayesian approach, which provides a complete picture of parametric uncertainty in the form of a non-parametric posterior density. From this, one can compute any relevant estimates, such as the MAP, mode, etc. We allow the two HUX parameters to vary based on physically reasonable ranges specified in Table~\ref{tab:uncertain-parameters}.

\noindent 
\begin{table}
\caption{The eleven uncertain continuous parameters in the PFSS$\to$WSA$\to$HUX model chain are modeled with uniform priors with physically meaningful ranges taken from previous parametric studies by~\cite[\S 2]{lee_2011_source_surface}, \cite[\S 2.5]{arden_2014_source_surface}, \cite[Eq. 9]{meadors_2020_particle_filter}, \cite[Table 1]{kumar_2022_parametric}, and \cite[Table 1]{riley_2015_parametric}.}
\centering
\begin{tabular}{c c c c c}
\textbf{Parameter} & \textbf{Model} & \textbf{Description} & \begin{tabular}{c} \textbf{Prior}\\\textbf{Range}\end{tabular}& \begin{tabular}{c} \textbf{Deterministic}\\\textbf{Value}\end{tabular}\\
\hline
$r_{\mathrm{SS}}$ [$R_{\mathrm{S}}$] & PFSS & source surface height & $[1.5, 4]$ & $2.5$ \\
\hline
$v_{0}$ $[\frac{\mathrm{km}}{\mathrm{s}}]$ & WSA & minimum velocity  & $[200, 400]$ & $250$\\
\hline
$v_{1}$ $[\frac{\mathrm{km}}{\mathrm{s}}]$ & WSA & maximum velocity  & $[550, 950]$ & $750$\\
\hline
$\alpha$  & WSA & numerical parameter & $[0.05, 0.5]$ & $0.1$\\
\hline
$\beta$ & WSA & numerical parameter  & $[1, 1.75]$ & $1$\\
\hline
$w$ [rad] & WSA & \begin{tabular}{c}the width solar wind ramps up\\from low- to high-speed flow\\{at coronal-hole boundaries}\end{tabular} & $[0.01, 0.4]$& $0.02$\\
\hline
$\gamma$ & WSA & numerical parameter &  $[0.06, 0.9]$ & $0.9$\\
\hline
$\delta$ & WSA & numerical parameter &  $[1, 5]$ & $1.75$\\
\hline
$\psi$ & WSA & numerical parameter  & $[3, 4]$ & $3$\\
\hline
$\alpha_{\mathrm{acc}}$ & HUX & acceleration factor & $[0, 0.5]$ & $0.15$\\
\hline
$r_{h}$ [$R_{\mathrm{S}}$] & HUX & 
\begin{tabular}{c} radial location at which\\the acceleration ends \end{tabular} & $[30, 60]$ & $50$
\end{tabular}
\label{tab:uncertain-parameters}
\end{table}

\subsection{Sobol’ Indices} \label{sec:sobol}
To introduce the notion of global sensitivity indices, let $(\Omega, \mathcal{F}, \mathbb{P})$ be a probability space with sample space $\Omega$, $\sigma$-algebra $\mathcal{F}$, and the probability measure $\mathbb{P}$, where $X: \Omega \to \mathcal{X}$ is a random vector with its entries being independent random variables $X_{i}$ for $i=1, \ldots d$. We denote with $x=X(\omega)$ a sample (realization) of the random vector $X$ for a given event $\omega \in \Omega$.
From the independence assumption, the joint probability density function (pdf) $\pi(x)$ is the product of the marginals, i.e. $\pi(x) = \pi_{1}(x_{1})\pi_{2}(x_{2}) \dotsb \pi_{d}(x_{d})$. We consider a generic model $f: \mathcal{X} \to \mathcal{Y}$ that maps a $d$-dimensional input parameter $x = [x_{1}, x_{2}, \ldots, x_{d}]^{\top} \in \mathcal{X} \subseteq \mathbb{R}^{d}$ to a scalar QoI $y \in \mathcal{Y} \subseteq \mathbb{R}$. We assume that $f$ is square-integrable with respect to $\pi$, such that the expectation (mean) $\mathbb{E}[f(X)] = \int_{\mathbb{R}^{d}} f(x) \pi(x) \mathrm{d}x$ and variance $\mathbb{V}\mathrm{ar}[{f(X)}] = \int_{\mathbb{R}^{d}} \left(f(x) - \mathbb{E}[f(x)]\right)^2 \pi(x) \mathrm{d}x$ of the QoI are both finite. 

\textbf{Definition 1 (Sobol' Indices)}
The \emph{first-order Sobol' sensitivity indices} measure the main variance contribution due to the $i$th random input parameter, such that
\begin{equation}\label{first-order-indices}
    S_{i} \coloneqq \frac{\mathbb{V}\mathrm{ar}_{X_{i}}[\mathbb{E}_{X_{\sim i}}(f(X)|X_{i})]}{\mathbb{V}\mathrm{ar}[f(X)]}, \qquad i=1, \ldots, d,
\end{equation}
where $\mathbb{V}\mathrm{ar}_{X_{i}}$ denotes the variance with respect to only the $X_{i}$ random input parameter, $\mathbb{V}\mathrm{ar}$ without subscript denotes variance involving all parameters, and $X_{\sim i}$ denotes all random input parameters but $X_{i}$.
The \emph{second-order Sobol' sensitivity indices} measure the secondary variance contribution due to the interaction of the $i$th and $j$th parameters (where $i\neq j$), such that
\begin{equation*}
    S_{ij} \coloneqq \frac{\mathbb{V}\mathrm{ar}_{X_{i}, X_{j}}[\mathbb{E}_{X_{\sim i, j}}(f(X)|X_{i}, X_{j})]}{\mathbb{V}\mathrm{ar}[f(X)]} - S_{i} - S_{j}.
\end{equation*}
The \emph{total-order Sobol' sensitivity indices} measure the total variance contributions of the $i$th parameter, such that
\begin{equation}\label{total-order-indices}
\begin{split}
    T_{i} &\coloneqq S_{i} + \sum_{j=1}^{d} S_{ij} + \mathrm{H.O.T.}= 1 - \frac{\mathbb{V}\mathrm{ar}_{X_{\sim i}}[\mathbb{E}_{X_{i}}(f(X)| X_{\sim i})]}{\mathbb{V}\mathrm{ar}[f(X)]} =  \frac{\mathbb{E}_{X_{\sim i}}[\mathbb{V}\mathrm{ar}_{X_{i}}(f(X)| X_{\sim i})]}{\mathbb{V}\mathrm{ar}[f(X)]},
\end{split}    
\end{equation}
where $\mathrm{H.O.T.}$ refers to higher-order terms. The first, second, and higher-order indices sum up to 1, such that 
\begin{equation*}
    \sum_{i=1}^{d} S_{i} + \sum_{i=1}^{d} \sum_{j=2, j>i}^{d} S_{ij} + \cdots+ S_{1 2\ldots d} =1.
\end{equation*}

Notice that if the total-order index $T_{i} \approx 0$, then $\mathbb{E}_{X_{\sim i}}[\mathbb{V}\mathrm{ar}_{X_{i}}(f(X)| X_{\sim i})] \approx 0$, which, by the non-negativity of the variance operator, implies that $\mathbb{V}\mathrm{ar}_{X_{i}}(f(X)| X_{\sim i}) \approx 0$. Therefore, if $T_{i} \approx 0$, the uncertainty in $X_{i}$ hardly influences the variance of the QoI, and $X_{i}$ can be deemed as non-influential.  

Sobol' sensitivity indices can not be computed in closed form except for QoIs that are integrable with respect to $\pi(x)$ (the joint probability of the uncertain parameters $X$). \ref{sec:appendix-a} shows that the sensitivity indices can be computed analytically for the simple Wang-Sheeley model~\cite{wang_sheeley_1990_ws}. However, the QoI that we consider (like most model QoIs arising from simulations of complex systems) is not integrable with respect to $\pi(x)$. Thus, we need to approximate the indices numerically.

\subsection{Estimating Sobol’ Indices via Monte Carlo Integration} \label{sec:mc-sobol}
The first- and total-order Sobol' sensitivity indices described in Eqns.~\eqref{first-order-indices} and \eqref{total-order-indices} can be estimated via Monte Carlo (MC) integration, which requires $N(d+2)$ model evaluations, where $N$ is the number of independent samples of $X$ and $d$ is the number of uncertain parameters. Since each model evaluation is independent of the other, the MC model evaluations can be easily computed in parallel. The four-step algorithm of~\cite{saltelli_estimator_2002}, which is based on~\cite{sobol_2001_theory} original work, is implemented as follows:
\begin{enumerate}
    \item Draw $2N$ quasi-random samples of the random vector $X$ and store them as
    \begin{equation*}
        A = \begin{bmatrix}
        x_{1}^{(1)} & \ldots & x_{d}^{(1)}\\
        \vdots & & \vdots\\
        x_{1}^{(N)} &\ldots &x_{d}^{(N)}
        \end{bmatrix} \in \mathbb{R}^{N \times d}
        \qquad \mathrm{and} \qquad
        B = \begin{bmatrix}
        x_{1}^{(N+1)} & \ldots & x_{d}^{(N+1)}\\
        \vdots & & \vdots\\
        x_{1}^{(2N)} &\ldots &x_{d}^{(2N)}
        \end{bmatrix}\in \mathbb{R}^{N \times d},
    \end{equation*}
    where $x_{i}^{(j)}$ denotes the $i$th entry and $j$th sample of $X$. Quasi-MC methods generate near-random samples that aim to distribute well over the parameter space. These sampling strategies usually result in a faster rate of convergence in MC integration. We use Latin hypercube sampling developed by~\cite{mckay_1979_lhs}. Other common quasi-random low-discrepancy sequences are Sobol'~\cite{sobol_1967_quasi_mc} and Halton~\cite{halton_1960_quasi_mc}.
    \item Define matrices $C^{(i)}$ for $i=1, 2, \ldots, d$, which are a copy of $B$ except the $i$th column is replaced by $A(:, i)$, the $i$th column of $A$, so that 
    \begin{equation*}
        C^{(i)} = \begin{bmatrix}
        \vert & &\vert& & \vert \\
        B(:, 1)& \ldots &A(:, i)&\ldots& B(:, d)\\
        \vert & &\vert & &\vert \\
        \end{bmatrix}\in \mathbb{R}^{N \times d}, \qquad i=1, \ldots, d.
    \end{equation*}
    \item Evaluate the QoI for each row of the matrices $A, B, C^{(i)}$, denoted as $A(j,:)$, $B(j,:)$, $C^{(i)}(j,:)$, i.e. 
    \begin{equation*}
        y_{A}^{(j)} = f(A(j,:)) \in \mathbb{R}, \qquad y_{B}^{(j)} = f(B(j,:)) \in \mathbb{R} \qquad \mathrm{and} \qquad y_{C^{(i)}}^{(j)} = f({C^{(i)}}(j,:))\in \mathbb{R},
    \end{equation*}
    for $j = 1, \ldots, N$. The evaluation of $y_{A}$ and $y_{B}$ requires $2N$ model evaluations, whereas the evaluation of $y_{C^{(i)}}$ requires $d \cdot N$ model evaluations, which results in a total of $N (d+2)$ model evaluations. 
    \item Estimate using MC integration the first-order $S_{i}$ and total-order $T_{i}$ sensitivity indices for $i=1, \ldots, d$. We use the unbiased Janon/Monod’s estimator~\cite{janon_estimator_2014, monod_estimator_2006}, such that
    \begin{align*}
        S_{i} &\approx \frac{\frac{1}{N} \sum_{j=1}^{N} y_{A}^{(j)} y_{C^{(i)}}^{(j)} - \left(\frac{1}{N} \sum_{j=1}^{N} y_{A}^{(j)}\right)\left(\frac{1}{N} \sum_{j=1}^{N} y_{C^{(i)}}^{(j)}\right) }{\frac{1}{N} \sum_{j=1}^{N}\left(y_{A}^{(j)}\right)^2 - \left(\frac{1}{N}\sum_{j=1}^{N}y_{A}^{(j)} \right)^{2}}, \\
        T_{i} &\approx 1 - \frac{\frac{1}{N}\sum_{j=1}^{N} y_{B}^{(j)} y_{C^{(i)}}^{(j)} -\left(\frac{1}{N} \sum_{j=1}^{N} \left(\frac{y_{B}^{(j)} + y_{C^{(i)}}^{(j)}}{2}\right)\right)^2 }{\frac{1}{2N} \sum_{j=1}^{N} \left(\left(y_{B}^{(j)}\right)^2 + \left(y_{C^{(i)}}^{(j)}\right)^{2} \right) - \left( \frac{1}{N} \sum_{j=1}^{N} \left(\frac{y_{B}^{(j)} + y_{C^{(i)}}^{(j)}}{2}\right)\right)^2}. 
    \end{align*}
\end{enumerate}

The algorithm intuition can be explained as follows. The first-order sensitivity estimation is based on the product of $y_{A}$ and $y_{C^{(i)}}$, which multiplies the QoI with input $A$ and the QoI with input $C^{(i)}$ where all parameters except $X_{i}$ have been re-sampled. Intuitively, if $X_{i}$ is influential then $y_{A}$ and $y_{C^{(i)}}$ are correlated and $S_{i}$ is large. We can intuit the estimation of the total-order indices $T_{i}$ in a similar way. The product of $y_{B}$ and $y_{C^{(i)}}$ multiplies the QoI with input $B$ and the QoI with input $C^{(i)}$ where we only re-sample $X_{i}$. Thus, if $X_{i}$ is influential then $y_{B}$ and $y_{C^{(i)}}$ are not correlated and $T_{i}$ is large.

There are many other MC Sobol' sensitivity indices estimators, see~\cite{puy_2022_gsa_comparison} for a comprehensive comparison. Saltelli's~\cite{saltelli_estimator_2002} and Jansen's~\cite{jansen_estimator_1999} estimators are also commonly used estimators. Such estimators require $N(d+2)$ model evaluations to compute the first and total-order indices. Computing second-order indices requires additional model evaluation, i.e. a total of $N(2d+2)$ model evaluations. We thus only compute first and total-order indices, which give a strong indication if a parameter is influential or not. We compared Saltelli's, Jansen's, and Janon/Monod's estimators and found that Janon/Monod’s estimator resulted in faster convergence for our study. 

\subsection{Global Sensitivity Analysis Numerical Results} \label{sec:gsa-results}
We perform global sensitivity analysis using the PFSS$\to$WSA$\to$HUX model chain for CR 2048 (September 21st, 2006 to October 18th, 2006), CR 2053 (February 4th, 2007 to March 4th, 2007), and CR 2058 (June 21st, 2007 to July 18th, 2007). The three CRs occurred during the declining phase of solar cycle 23. The eleven model input parameters are listed in Table~\ref{tab:uncertain-parameters} and are described in Section~\ref{sec:uncertain-parameters}. We use $N=10^{4}$ Latin Hypercube samples~\cite{mckay_1979_lhs} to estimate the Sobol' sensitivity indices via MC integration (Section~\ref{sec:mc-sobol}), which requires $N(d+2) = 1.3 \times 10^{5}$ model evaluations for each CR. We consider two QoIs: the root mean squared error (RMSE) between ACE velocity measurements and the model predictions at L1 and longitude-dependent model predictions at L1 (independent of ACE observations). The results are discussed in the following subsections. 

\subsubsection{Sobol' Indices for RMSE} \label{sec:error-sobol-indicies}
\begin{figure}
    \centering
    \includegraphics[width=0.8\textwidth]{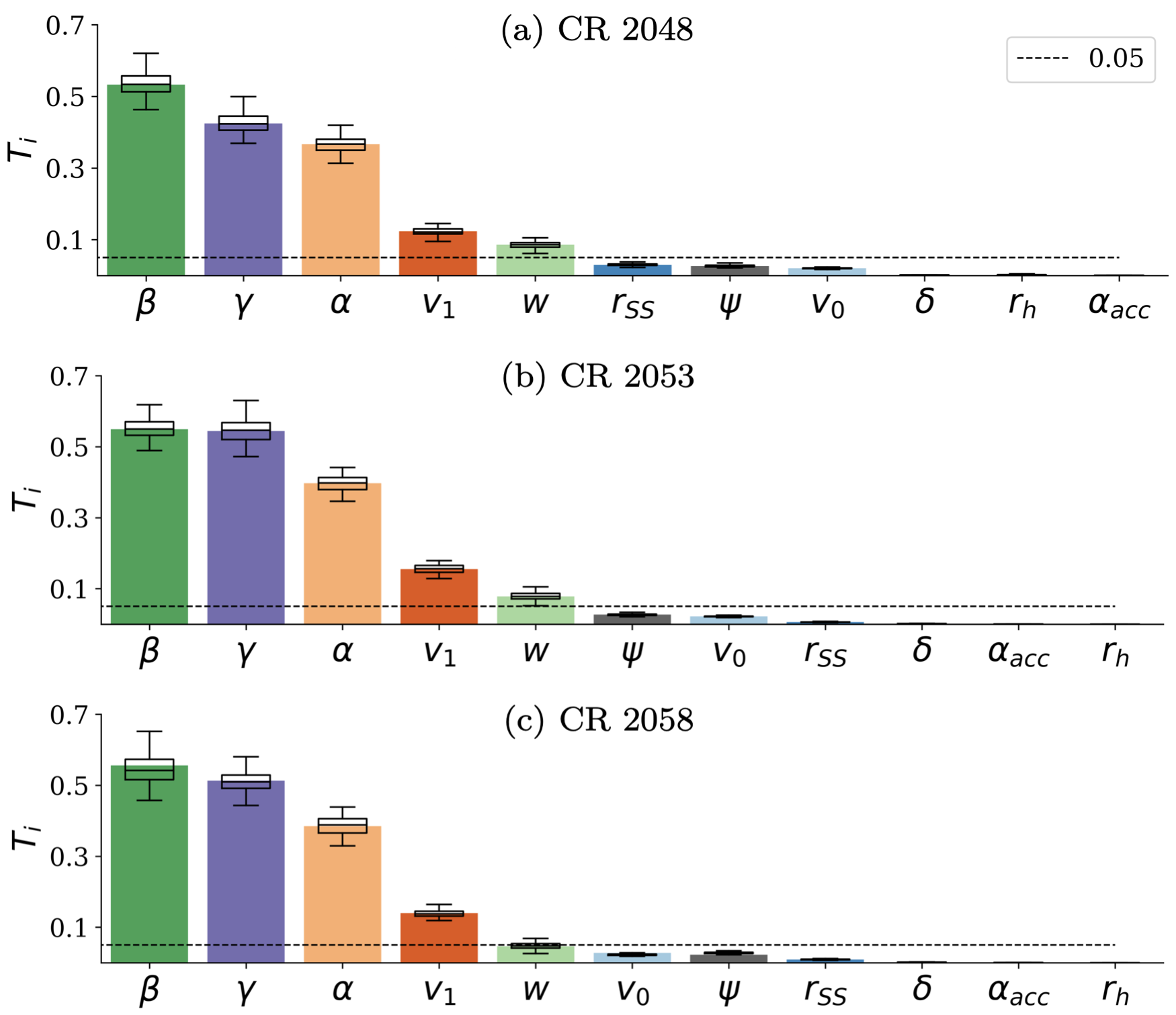}
    \caption{The total-order indices $T_{i}$ of the RMSE between the model chain and ACE observations are shown for (a) CR 2048, (b) CR 2053, and (c) CR 2058. The box plot for each index shows the uncertainty in the index estimate using bootstrapping with $N=3\times 10^{3}$ samples and $100$ replications. The box plots display the range between the first and third quartiles, with a middle line indicating the median. The whiskers represent the span from the minimum to maximum estimates. The results show that $r_{\mathrm{SS}}, \psi, \delta, v_{0}, r_{h}, \alpha_{\mathrm{acc}}$ are non-influential as their total-order indices are lower than 0.05 (shown in dashed black horizontal line).}
    \label{fig:RMSE-total-indicies-ordered}
\end{figure}

The QoI $f(X)$ for which we compute the Sobol' sensitivity indices is the RMSE between the model chain solar wind radial velocity prediction at L1 and ACE at 1-hour cadence observations. 
Figure~\ref{fig:RMSE-total-indicies-ordered} shows the total-order indices for CR 2048, CR 2053, and CR 2058. The total-order indices for all three CRs have the same ordering for the five most influential parameters, i.e. $\beta, \gamma, \alpha, v_{1}, w$ (listed in descending order). The five most influential parameters are all WSA model parameters. 
The other six parameters $r_{\mathrm{SS}}, \psi, v_{0}, \delta, r_{h}, \alpha_{\mathrm{acc}}$ are deemed as non-influential since their total-order indices are less than 0.05. 
We also estimate the uncertainty in the estimated total-order indices by using bootstrapping with $N=3 \times 10^{3}$ samples and $100$ replications. The box plots for each index are shown in Figure~\ref{fig:RMSE-total-indicies-ordered}. The uncertainty in the estimated total-order indices does not influence the classification between influential and non-influential parameters. 
The total-order indices show that six out of the eleven uncertain parameters are non-influential, and subsequently, they hardly contribute to the predicted solar wind radial velocity variability at L1. We, therefore, set the six non-influential parameters to their fixed deterministic values (see Table~\ref{tab:uncertain-parameters}) in the subsequent Bayesian inference, which facilitates \textit{a posteriori} dimensionality reduction and significant computational speed up for performing MCMC. 

\begin{figure}
    \centering
    \includegraphics[width=\textwidth]{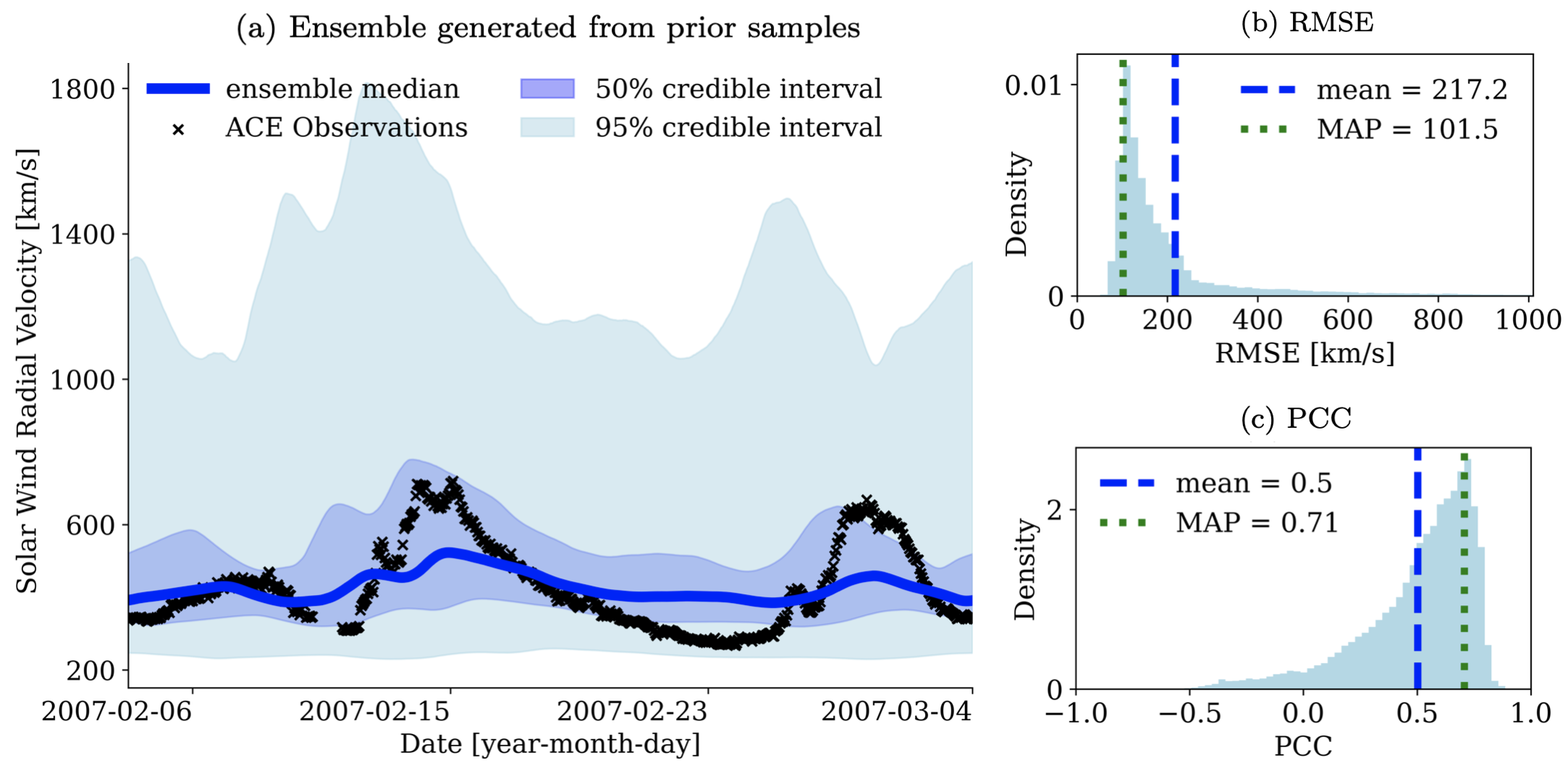}
    \caption{(a) Ensemble generated from prior samples of the global sensitivity analysis model evaluations for CR 2053. The credible interval shows that parametric uncertainty in the model chain results in very high uncertainty in the solar wind radial velocity predictions at L1. A histogram of the ensemble RMSE and PCC are shown in (b) and (c), respectively. }
    \label{fig:GSA-samples-CR2053}
\end{figure}

Figure~\ref{fig:GSA-samples-CR2053}(a) shows an ensemble of the global sensitivity analysis model evaluations for CR 2053. We plot the median and 50\% and 95\% credible interval of the $1.3 \times 10^{5}$ model evaluations used to compute the sensitivity indices (which were constructed via prior density samples). A \emph{credible interval} is an interval within which the ensemble members fall with a particular probability. The 95\% credible interval shown in Figure~\ref{fig:GSA-samples-CR2053}(a) spans an excessively large range and includes non-physical solutions (for example, solar wind radial velocity at $1800\frac{\mathrm{km}}{\mathrm{s}}$). This is because the uncertainties in the model chain parameters highly influence the solar wind velocity predictions at L1. We aim to reduce such large parametric uncertainties via Bayesian inference; see Section~\ref{sec:bayes-and-mcmc}. Figure~\ref{fig:GSA-samples-CR2053}(b) and Figure~\ref{fig:GSA-samples-CR2053}(c) show histograms of the RMSE and Pearson correlation coefficient (PCC) between the global sensitivity analysis simulations in comparison to ACE observations for CR 2053. The histograms show that the RMSE mean is $217.2 \frac{\mathrm{km}}{\mathrm{s}}$ and the PCC mean is 0.5.  Also, the RMSE MAP is $101.5 \frac{\mathrm{km}}{\mathrm{s}}$ and the PCC MAP is 0.71. By reducing the uncertainty in the model parameters, we expect the ensemble to be more accurate~\cite[\S 8.1]{ralph_smith_uq_2013}. 

\subsubsection{Longitude-Dependent Sobol' Indices} \label{sec:time-dependent-sobol-indicies}
\begin{figure}
    \centering
    \includegraphics[width=\textwidth]{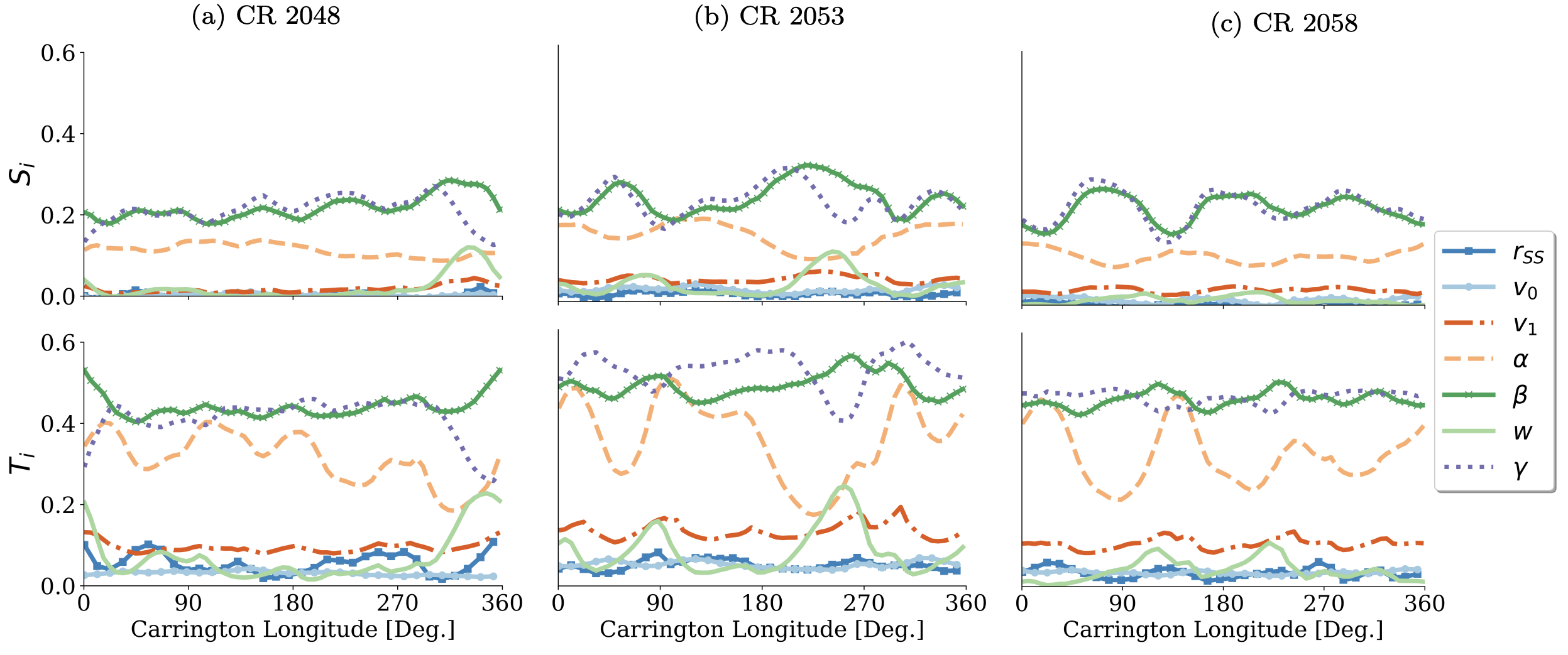}
    \caption{Longitude-dependent (top) first-order $S_{i}$ and (bottom) total-order indices $T_{i}$ for (a) CR 2048, (b) CR 2053, and (c) CR 2058. We do not plot the indices of $\psi, \delta, \alpha_{\mathrm{acc}}, r_{h}$ since their maximum first-order and total-order indices (in longitude) are less than 0.05. The parameters $\beta, \gamma, \alpha, v_{1}$ are the most influential across all longitudinal locations, whereas $w$ seems to be more longitudinal (or time) dependent. For example, during CR 2048, $w$ is only influential from approximately $280^{\circ}$ to $360^{\circ}$ in longitude.}
    \label{fig:time-dependent-indicies}
\end{figure}

We define longitude-dependent (or time-dependent) QoI, which is the solar wind radial velocity predictions at L1 at a 1-hour cadence. In contrast to the RMSE indices, defined in Subsection~\ref{sec:error-sobol-indicies}, the longitude-dependent QoI is independent of ACE observations. Figure~\ref{fig:time-dependent-indicies} presents the seven largest first-order $S_{i}$ and total-order $T_{i}$ indices as a function of Carrington longitude for CR 2048, CR 2053, and CR 2058. We do not plot the first- and total-order indices of $\psi, \delta, \alpha_{\mathrm{acc}}, r_{h}$ since their maximum indices (in longitude) are less than 0.05. The five most influential parameters during all three CRs are $\beta, \gamma, \alpha, v_{1}, w$, which agree with the RMSE indices results, see Figure~\ref{fig:RMSE-total-indicies-ordered}. The first-order indices are significantly smaller than the total-order indices, which indicates that higher-order interactions between parameters influence the predicted solar wind velocity variability. Additionally, the influence of $w$, the width over which the solar wind ramps up from low- to high-speed flow at coronal-hole boundaries, is mainly around a small longitudinal region. For example, during CR 2053, $w$ is only influential from approximately 190$^{\circ}$ to 280$^{\circ}$. We suspect this is because ACE's footpoints lie closer to the center of a low-latitude coronal hole at approximately $250^{\circ}$ to $310^{\circ}$, see Figure~\ref{fig:model-components-CR2053}(d). The large distance to coronal hole boundary $d$ corresponds to high solar wind speed in this region, which is then advected to 190$^{\circ}$ to 280$^{\circ}$ at L1 (at $\frac{1}{v}$ speed, see Eq.~\eqref{hux-underlying-equation-PDE}). Thus, $w$ seems to be influential only in regions where $d$, the distance to the coronal hole boundary, in the WSA relation is large. 
Similarly, the influence of $\alpha$ seems to be correlated to $f_{p}$, e.g. when $f_{p}$ is low at approximately $220^{\circ}$ to $300^{\circ}$ in longitude during CR 2053, see Figure~\ref{fig:model-components-CR2053}(c), we see a decrease in the total-order index of $\alpha$, see Figure~\ref{fig:time-dependent-indicies}(b).

\section{Bayesian Inference via Markov Chain Monte Carlo Sampling} \label{sec:bayes-and-mcmc}
After identifying the set of influential parameters via variance-based global sensitivity analysis, our goal is to learn the uncertainties of such influential parameters, which we achieve through Bayesian inference. Bayesian parameter estimation leverages Bayes' theorem to learn the pdf of uncertain model parameters given observational data. 
Samples from such pdfs can be directly obtained using Markov chain Monte Carlo (MCMC) algorithms. These samples are then used to generate an ensemble prediction to quantify and reduce the effect of the parametric uncertainty on the QoI. The following subsections introduce Bayesian inference and MCMC sampling.  

\subsection{Bayesian Parameter Estimation}\label{sec:bayes}
The philosophy behind Bayesian statistics is that the model parameters are random variables with an unknown pdf. This differs from the frequentist perspective, where the parameters are assumed deterministic but unknown. 
In the Bayesian setting, we seek to estimate the pdf of model influential parameters $X$ given a parameter-dependent QoI $f(\cdot)$ (e.g. solar wind radial velocity at L1) and measurements of the QoI $z = \{z_{1}, z_{2}, \ldots, z_{n}\}$ (e.g. ACE radial velocity measurements) taken at time instances $t_{1} < t_{2} < \ldots < t_{n}$ (e.g., at a 1-hour cadence). In other words, we aim to estimate the conditional pdf $\pi(x|z)$, which is referred to as the \textit{posterior density} or simply \emph{posterior}. 
The posterior density can be evaluated via Bayes' rule:
\begin{equation*}
    \pi(x|z) = \frac{\pi(z|x)\pi(x)}{\pi(z)} = \frac{\pi(z|x)\pi(x)}{\int_{\mathbb{R}^{d}} \pi(z|x)\pi(x)\mathrm{d}x} \propto \pi(z|x)\pi(x),
\end{equation*}    
where $\pi(x)$ is the \emph{prior}, $\pi(z|x)$ is the \emph{likelihood}, and $\pi(z)$ is the \emph{evidence} (also referred to as the marginal likelihood or normalizing constant). The parameters $x$ are samples of the random variable $X$, and the observations $z_{i}$ are samples of the random variable $Z_{i}$. Most often, the evidence can not be properly defined, so we estimate the posterior up to a normalizing constant. The posterior density can be continuously refined as more measurements are included.

The prior density $\pi(x)$ is chosen based on physically meaningful ranges and previous studies in the literature; see Table~\ref{tab:uncertain-parameters} for the list of the uniform prior densities used in this study. Generally, the priors are not restricted to uniform densities and may weigh favorable values more heavily. However, if prior knowledge is of questionable accuracy, it is better to use non-informative priors~\cite[\S 8.1]{ralph_smith_uq_2013}.

We assume that the QoI of the model and measurements are related via 
\begin{equation}\label{gaussian-error-assumption}
    Z_{i} = f(X; t_{i}) + \epsilon_{i}, \qquad i = 1, \ldots, n,
\end{equation}
where $Z_{i}$ is a random variable representing the measurements at time instance $t_{i}$, $f(X; t_{i})$ is the QoI at time instance $t_{i}$ and $\epsilon_{i}$ is a random variable representing the discrepancies between the QoI and measurements. Here, we model $\epsilon_{i}$ as a Gaussian random variable with zero mean and standard deviation $\sigma \in \mathbb{R}_{+}$. We note that the model discrepancies and measurement noise are modeled as additive and mutually independent of $X$. Thus, by the assumption of Gaussian additive error and independence of measurements, we can write the likelihood as
\begin{equation*}
    \pi(z|x) = \prod_{i=1}^{n} \frac{1}{\sqrt{2\pi\sigma^{2}}} \exp \left(-\frac{1}{2\sigma^{2}} \left[z_{i} - f(x; t_{i})\right]^{2} \right) \propto \exp \left(-\frac{1}{2\sigma^{2}} \sum_{i=1}^{n} \left[z_{i} - f(x; t_{i}) \right]^{2} \right).
\end{equation*}
Although we derived an expression for the prior and likelihood densities, we can not directly sample from the posterior density since the evidence (or normalizing constant) remains unknown. To overcome this issue, MCMC algorithms enable sampling from arbitrary pdfs and allow for the unbiased estimation of the posterior density, mean, and variance. 

\subsection{Markov Chain Monte Carlo Sampling}\label{sec:mcmc-sampling}
MCMC algorithms generate samples from an arbitrary target pdf (such as posterior pdf) by generating a random walk in the parameter space that draws a representative set of samples from the target pdf. The random walk is a Markov chain, with the property that each sample only depends on the position of the previous sample. MCMC algorithms converge to the exact target pdf as the number of samples increases. This convergence property is established by the ergodicity property of MCMC, which requires the Markov chain to be aperiodic, irreducible, and reversible with respect to the target pdf~\cite{rosenthal_2004_mcmc}. 

The first and most frequently used MCMC algorithm is the Metropolis-Hastings algorithm~\cite{metropolis_1953_mcmc, hastings_1970_mcmc} developed at Los Alamos National Laboratory. The Metropolis–Hastings algorithm generates samples from an arbitrary pdf iteratively. The samples are drawn from a proposal density which is chosen \textit{a priori} and depends on the position of the previous sample of the Markov chain. A proposed sample is then accepted or rejected with some probability. If accepted, the proposed sample is appended to the Markov chain and used to generate the next sample; otherwise, if rejected, the proposed sample is discarded, and the previous sample is appended to the Markov chain. A common choice of proposal density is the Gaussian distribution centered at the previous sample location. 

We employ the affine invariant ensemble sampler~(AIES) developed by~\cite{goodman_2010_emcee}, which we describe in detail in~\ref{sec:aies}. The AIES offers several advantages over the Metropolis-Hastings algorithm. Firstly, AIES has only two hyperparameters, while Metropolis-Hastings has approximately $d^2$ hyperparameters, where $d$ represents the number of uncertain parameters. Secondly, AIES remains invariant to affine transformations, enabling easy sampling from anisotropic pdfs. Finally, AIES can be run in parallel, leading to considerably faster convergence (measured by the integrated autocorrelation time, see~\ref{sec:aies}) compared to the single-chain Metropolis-Hastings algorithm. We also elaborate in \ref{sec:aies} on MCMC burn-in and convergence assessment which are important heuristics that verify the Markov chain has reached a stationary distribution.

\subsection{Markov Chain Monte Carlo Numerical Results} \label{sec:mcmc-results}
We use the AIES sampler (described in~\ref{sec:aies}) to uncover the posterior density of the five most influential parameters $\beta, \gamma, \alpha, v_{1}, w$, for CR 2048 to CR 2058 (spanning from September 21st, 2006 to July 18th, 2007). We performed the computationally expensive global sensitivity analysis for CR 2048, CR 2053, and CR 2058, and assume the results hold for the whole time period between CR 2048 to CR 2058. We exclude CR 2051 since, during this time period, three CMEs reached L1, and the PFSS$\to$WSA$\to$HUX model chain does not account for transient events such as CMEs.
We assume in Eq.~\eqref{gaussian-error-assumption} that the model chain solar wind radial velocity predictions at L1 and ACE measurements (at 1-hour cadence) are related via Gaussian error with mean zero and standard deviation $\sigma = 80 \frac{\mathrm{km}}{\mathrm{s}}$. The standard deviation is chosen from previous parametric studies by \cite[Table 1]{reiss_2020_parametric} and \cite[Figure 8]{kumar_2022_parametric}.
The posteriors are approximated with $2.6 \times 10^{4}$ MCMC iterations, $10^{3}$ iterations excluded for burn-in, and $L=250$ walkers, resulting in $M=6.25 \times 10^{6}$ MCMC posterior samples per CR. 

\subsubsection{Markov Chain Monte Carlo Posterior Densities}
\begin{figure}
    \centering
    \includegraphics[width=0.95\textwidth]{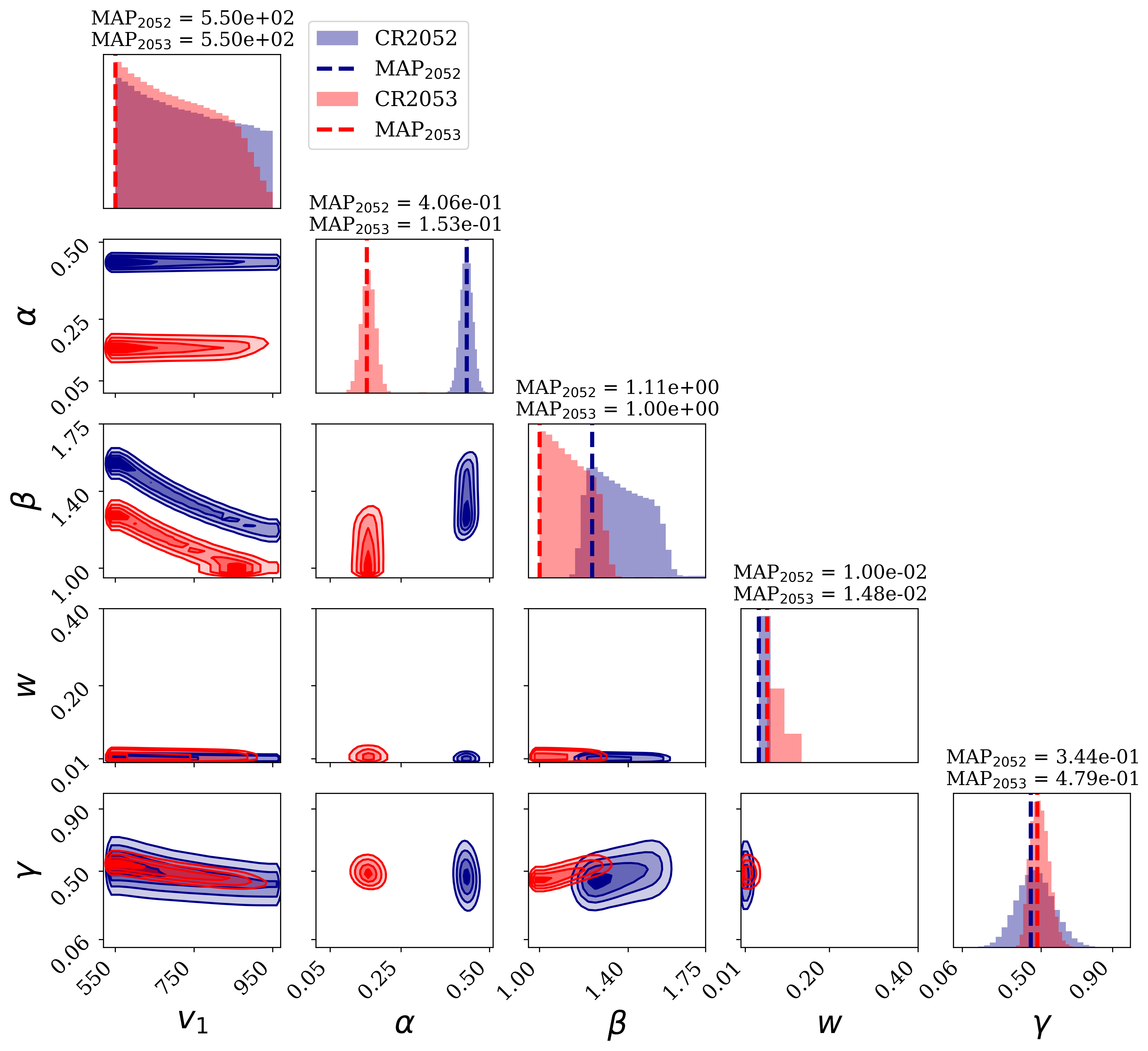}
    \caption{A corner plot of the posterior density of the five most influential parameters $v_{1}, \alpha, \beta, w, \gamma$ during CR 2052 (blue) and CR 2053 (red). The corner plot shows the MCMC samples in two-dimensional and one-dimensional projected parameter space. The dashed line shows the estimated MAP of each parameter. }
    \label{fig:corner-plot-CR2052-CR2053}
\end{figure}

The posterior densities for CR 2052 and CR 2053 in one- and two-dimensional projected parameter space are shown in Figure~\ref{fig:corner-plot-CR2052-CR2053}. 
It is apparent that the marginal posterior of $v_{1}$ is uniformly distributed (resembling the prior density in Table~\ref{tab:uncertain-parameters}), meaning that $v_{1}$ is highly uncertain and can take any value in the prior range with equal probability. This means that the likelihood function is flat in the $v_{1}$ direction, i.e. $v_{1}$ may not be identifiable from ACE observations. The corner plot shows that $\beta$ and $v_{1}$ are negatively correlated. Note that the marginal posteriors of parameters $\alpha$ and $\beta$ have little to no support overlap in CR 2052 and CR 2053. This suggests that such parameters are difficult to predict in advance. 

\begin{figure}
    \centering
    \includegraphics[width=\textwidth]{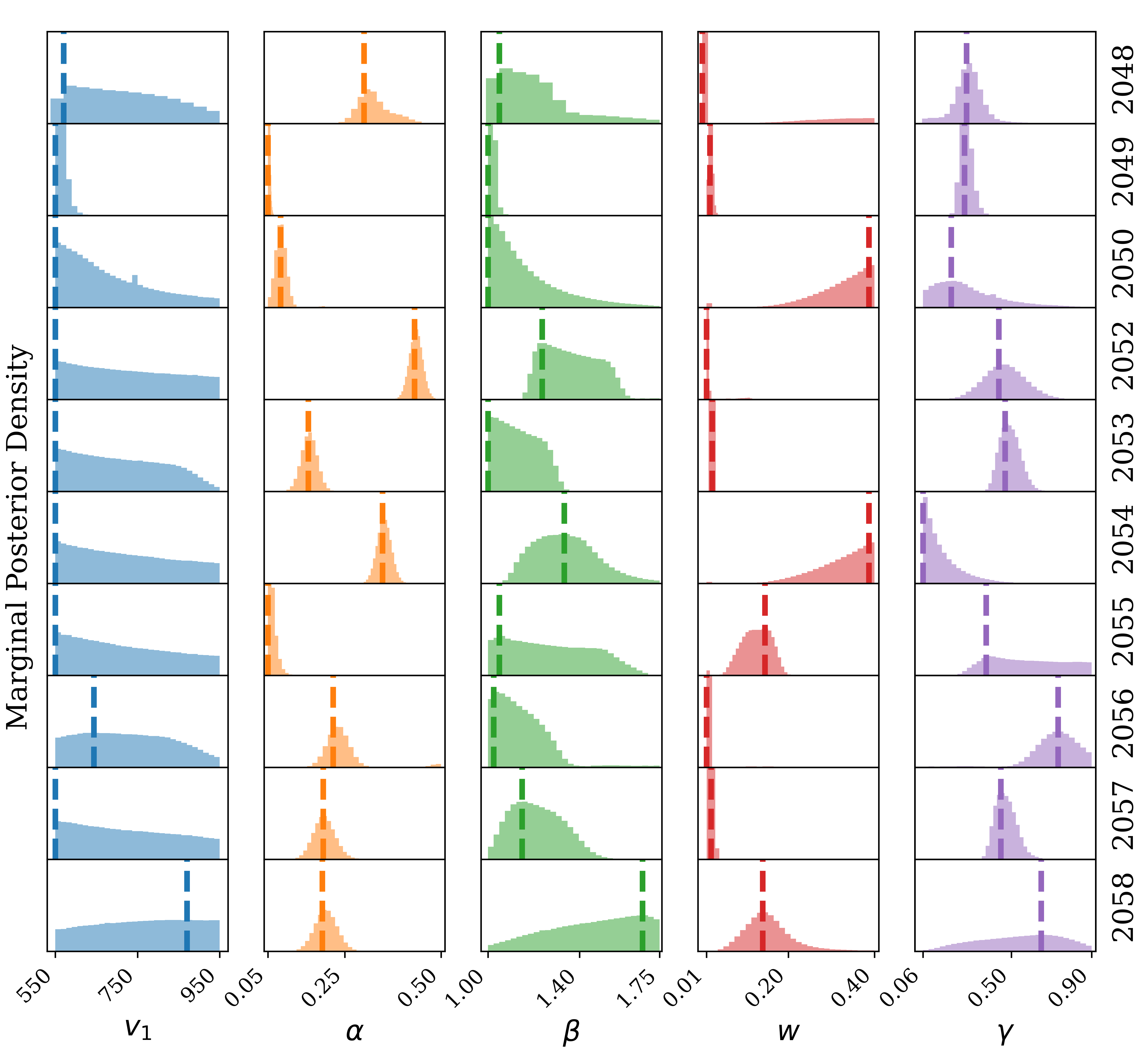}
    \caption{The marginal posterior densities of the five most influential parameters $v_{1}, \alpha, \beta, w, \gamma$ from CR 2048 to CR 2058 (excluding CR 2051). The solid line shows the estimated MAP of each parameter. The marginal posteriors change substantially from one CR to the next.}
    \label{fig:posterior-CR2048-to-CR2058}
\end{figure}

The marginal posterior densities for CR 2048 to CR 2058 (excluding CR 2051) are shown in Figure~\ref{fig:posterior-CR2048-to-CR2058}, which indicates that the posterior densities evolve from one CR to the next in a non-predictable fashion. For example, the marginalized posterior of $\alpha$ has relatively small support that varies randomly from one CR to the next. We also notice that the MAP, shown in dashed vertical lines, changes greatly from one CR to the next, which agrees with the previous parametric studies by~\cite{kumar_2022_parametric} and~\cite{riley_2015_parametric}. Since the posterior densities vary greatly from one CR to the next, it is not possible to use the posterior samples from a given CR to create an accurate ensemble prediction of the next CR (in contrast to the adaptive-WSA method proposed by~\cite{reiss_2020_parametric}). 
If the model chain is used for re-analysis studies, we recommend using the proposed UQ framework to generate accurate ensembles. The ensembles generated from the MCMC posterior samples will be highly accurate as the parameter posteriors are learned using observational data at L1 and will also be able to successfully capture the parametric uncertainty on the predictions. The next subsection discusses our ensemble prediction numerical results. 

\subsubsection{Ensemble Prediction} \label{sec:Ensemble-results}
\begin{figure}
    \centering
    \includegraphics[width=\textwidth]{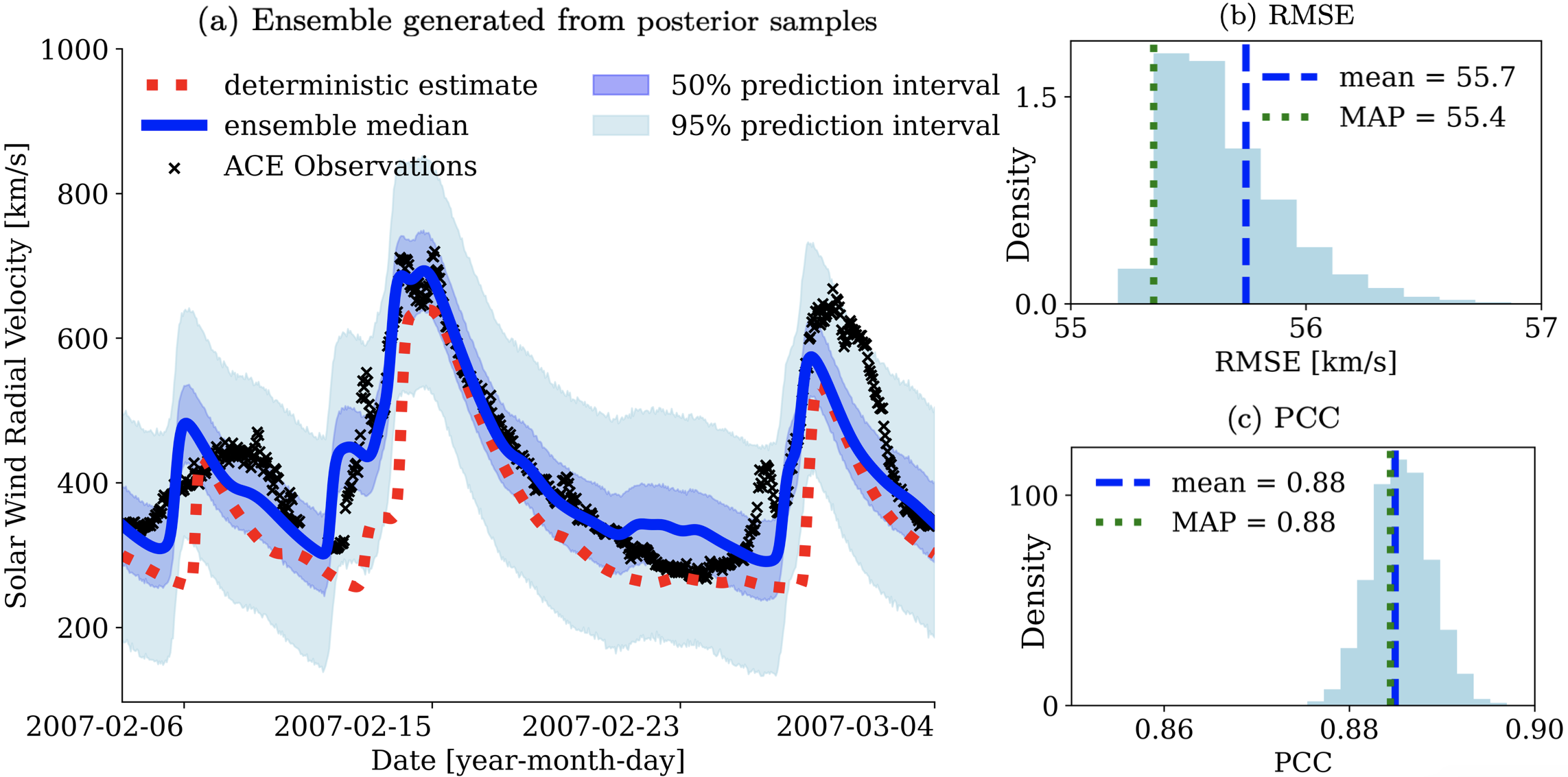}
    \caption{(a) Ensemble prediction with $5 \times 10^{3}$ ensemble members generated from posterior samples for CR 2053. The figure shows the ensemble statistics (median and prediction intervals), a model chain prediction when all parameters are set to their deterministic values (from Table~\ref{tab:uncertain-parameters}), which we label as `deterministic estimate', and ACE observations. Figures (b) and (c) show the RMSE and PCC of the ensemble in comparison to ACE observations, respectively.}
    \label{fig:ensemble-CR2053}
\end{figure}

\begin{figure}
    \centering
    \includegraphics[width=\textwidth]{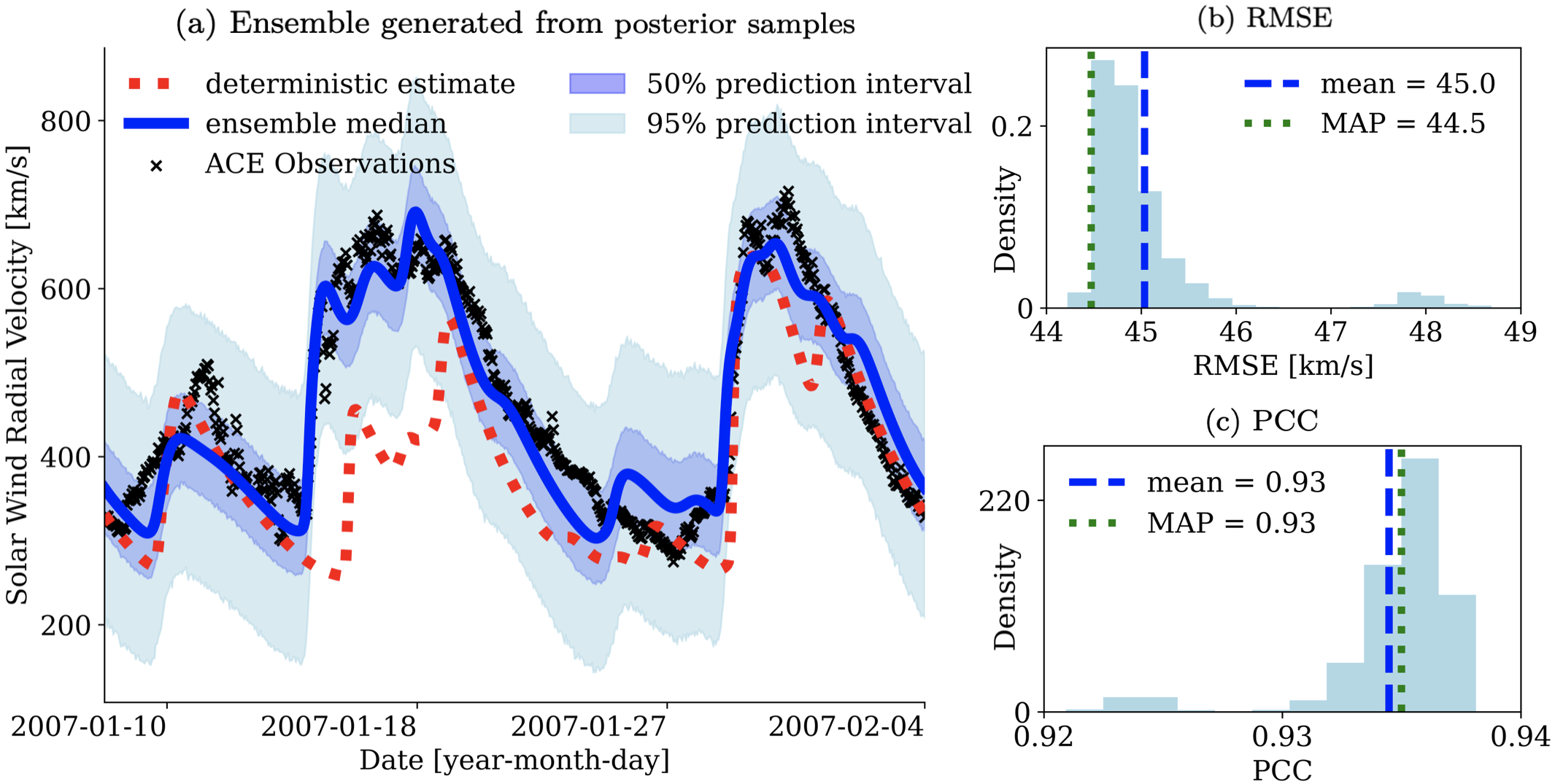}
    \caption{Same as Figure~\ref{fig:ensemble-CR2053} for CR 2052.}
    \label{fig:ensemble-CR2052}
\end{figure}

We generate an ensemble prediction based on varying model parameters using MCMC posterior samples.
The ensemble members are then used to compute ensemble statistics, such as the ensemble median and prediction interval. The \emph{prediction interval} accounts for both the propagated parametric uncertainty and assumed model discrepancy errors~\cite[\S 9.4.3]{ralph_smith_uq_2013}. 
The $(1 - \alpha) \times 100\%$ prediction interval for a fixed but unknown new observation $Z_{i}$ at time instance $t_{i}$ is the interval $[Z_{l}, Z_{u}]$ such that 
\begin{equation*}
    \mathbb{P}(Z_{l} \leq Z_{i} \leq Z_{u}) = 1 - \alpha,
\end{equation*}
where $Z_{i}$ is independent of the data used to construct the random variables $Z_{l}$ and $Z_{u}$~\cite[\S 9.4.1]{ralph_smith_uq_2013}. We estimate the interval $[Z_{l}, Z_{u}]$ via computing the $\alpha/2$th and $1 - \alpha/2$th quantiles of the set of ensemble members with added Gaussian model discrepancy errors. In general, one should be cautious when assessing the ensemble prediction via quantiles when the posterior predictive density, defined as $\pi(Z_{i}|Z)$, is multimodal. We have checked that the computed posterior predictive density is indeed unimodal for CR 2052 and CR 2053 (not shown here).
Figure~\ref{fig:ensemble-CR2053}(a) and Figure~\ref{fig:ensemble-CR2052}(a) show the median and 50\% and 95\% prediction interval of the $5\times 10^{3}$ ensemble members during CR 2053 and CR 2052, respectively. The ensemble is generated using posterior samples trained separately on each CR. 
Figure~\ref{fig:ensemble-CR2053}(b) and Figure~\ref{fig:ensemble-CR2052}(b) show a histogram of the RMSE of the ensemble members for each CR. Figure~\ref{fig:ensemble-CR2053}(c) and Figure~\ref{fig:ensemble-CR2052}(c) show a histogram of the PCC of the ensemble members for each CR. By comparing Figures~\ref{fig:GSA-samples-CR2053}(b/c) to Figures~\ref{fig:ensemble-CR2053}(b/c), it is apparent that the ensemble generated from the posterior density is able to substantially reduce the parametric uncertainty and improve the accuracy of the ensemble prediction. Specifically, the mean RMSE is reduced from $217.2 \frac{\mathrm{km}}{\mathrm{s}}$ to $55.7 \frac{\mathrm{km}}{\mathrm{s}}$, and the mean PCC is increased from 0.5 to 0.88. 
The results show that the proposed UQ framework is able to successfully reduce the uncertainty of the model parameters on the solar wind radial velocity prediction and should be utilized in further re-analysis studies.

Figure~\ref{fig:ensemble-CR2053}(a) and Figure~\ref{fig:ensemble-CR2052}(a) also show the model prediction with all parameters set to their deterministic values listed in Table~\ref{tab:uncertain-parameters}, which we label as `deterministic estimate'. The figures show that in both time periods, CR2052 and CR2053, the ensemble prediction after UQ is substantially more accurate than the deterministic estimate before UQ. In particular, the deterministic estimate underestimated the solar wind velocity in both time periods. For a more quantitative comparison, Table~\ref{tab:ensemble-vs-deterministic} lists the RMSE and PCC of ACE observations in comparison to the ensemble mean and deterministic estimate during CR2052 and CR2053. The numerical results show that the ensemble mean RMSE is nearly 50\% smaller than the deterministic estimate RMSE for both time periods. Thus, the proposed UQ framework is also able to substantially improve the accuracy of the model's solar wind radial velocity predictions at L1. 

\begin{table}
\caption{The RMSE and PCC of ACE observations in comparison to the model chain prediction with all parameters set to their deterministic values before UQ~(see Table~\ref{tab:uncertain-parameters}) vs. the ensemble mean after UQ during CR2052 and CR2053.}
\centering
\begin{tabular}{ l  c c c }
  \multicolumn{2}{l}{}  & RMSE ($\frac{\mathrm{km}}{\mathrm{s}}$) & PCC \\
  \hline
  \multirow{2}*{CR2052}  & Ensemble mean & 45.0 & 0.93 \\
       & Deterministic estimate & 115.6 & 0.73\\
  \hline
  \multirow{2}*{CR2053}  & Ensemble mean & 55.7 & 0.88 \\
       & Deterministic estimate & 97.7 & 0.86 \\
\end{tabular}
\label{tab:ensemble-vs-deterministic}
\end{table}
\section{Conclusions and Discussion} \label{sec:conclusion} 
The PFSS$\to$WSA$\to$HUX model chain is commonly used to predict the ambient solar wind radial velocity near Earth. The model chain has eleven uncertain input parameters that can not be directly measured since they are mainly non-physical. We, therefore, propose a comprehensive UQ framework for quantifying and reducing the parametric uncertainty of the model chain. The proposed framework utilizes variance-based global sensitivity analysis to reduce the dimensionality of the parameter space, followed by Bayesian inference to learn the full parameter pdfs via MCMC. We apply the UQ framework on a time period spanning from CR 2048 to CR 2058 during the declining phase of solar cycle 23. The sensitivity analysis results show that $\beta, \gamma, \alpha, v_{1}, w$ are the five most influential parameters in the model chain. These parameters are all WSA parameters. We learned the posterior densities of the five most influential parameters using AIES (an MCMC sampler). The posterior samples are then used to generate an ensemble prediction and quantify the parametric uncertainty in the predicted solar wind velocity. We found that the ensemble results are able to accurately quantify the uncertainty in the predictions and thus suggest the proposed UQ framework should be utilized in further re-analysis studies employing the model chain.

The Bayesian inference numerical results also show that the posterior densities vary randomly from one CR to the next. This is mainly due to the following reasons: (1) the model chain is not robust to the choice of WSA numerical parameters, and (2) the WSA model is overparameterized (i.e. needs to be reformulated for forecasting purposes). The reformulation of the WSA model will involve searching for a parsimonious model that is robust to its choice of parameters. Candidates of models that balance the trade-off between accuracy and parsimony can be found using sparse regression techniques with different regularization~\cite{brunton_2016_sindy}. The optimal model can later be selected via the Bayesian or Akaike information criterion~\cite{schwarz_1978_bic,akaike_1974_aic}. The substantial and unpredictable change in the posterior densities from one CR to the next questions the applicability of the model chain in operational real-time forecasting.

We suspect the drastic changes in the posterior densities are also due to the parameters trying to overcompensate the intrinsic and ``hard-wired'' limitations of each of the models (i.e. biases due to epistemic uncertainties). We next discuss such limitations. 
First, we do not have an accurate estimate of the photospheric fields~\cite{poduval_2020_photpsheric}. There are differences between magnetograms from different observatories. There are also different saturation levels and noise~\cite{riley_2014_magnetograms}. 
Second, the PFSS solutions rely on the existence of a spherical source surface, which does not exist~\cite{riley_2006_pfss_vs_mhd}. The sensitivity analysis results show that the choice of the source surface height is non-influential on the predicted solar wind velocity at L1, yet in the analysis we assume it exists. Also, the fields are not potential, particularly around active regions. 
Third, the WSA model has known inaccuracies, e.g. the expansion factor in the vicinity of pseudostreamers~\cite{riley_2015_parametric}, as well as unknown inaccuracies. 
Fourth, the HUX model assumes only radial propagation and neglects external forces and the pressure gradient~\cite{riley_HUXP1_2011}.
Fifth, time dependence is not included in synoptic maps and all throughout the model chain.
Thus, the physics simplifications in the model chain introduce model discrepancies between the spacecraft observations and model predictions. We assume such discrepancies are Gaussian distributed in the Bayesian inference setup. This is generally a reasonable assumption, which is necessary in order to formulate the likelihood in the Bayesian setting, yet it is important to point out that the model chain discrepancies are structured and are not i.i.d.

Future studies can incorporate the proposed UQ framework for learning the posterior densities of uncertain parameters for various (and more complex) space weather models, for example, the WSA-Enlil model~\cite{parsons_operational_wsa_enlil_2011}. It will be interesting to apply the proposed UQ framework to WSA-Enlil to make sure the WSA posteriors do not change drastically in time and verify that the WSA-Enlil model is reliable for real-time forecasting. Depending on the computational resources and computational complexity of the model at hand, one might need to incorporate surrogate models (computationally efficient approximate models), such as projection-based reduced-order models~\cite{benner_2015_prom, issan_sopinf_2022} and interpolatory surrogates~\cite{xiu_2002_gpce}, to compute Sobol' sensitivity indices and run MCMC. If the model is computationally efficient (i.e. order of seconds/minutes) we recommend using the MC methods presented in this study as they are unbiased estimators. Other unbiased estimators include multi-fidelity estimators, see~\cite{peherstorfer_2018_mf} for a detailed survey.

\appendix
\section{Global Sensitivity Analysis of the Wang-Sheeley Model: Analytic Results}\label{sec:appendix-a}
The Wang-Sheeley (WS) semi-empirical model developed by~\cite{wang_sheeley_1990_ws} is based on the inverse relationship between the solar wind speed and the magnetic field expansion factor $f_{p}$ (defined in Eq.~\eqref{expansion_factor}). The WS model relation is given by
\begin{equation*}
    v_{\mathrm{ws}}(f_{p}, v_0, v_1, \alpha) = v_{0} + \frac{v_{1} - v_{0}}{f_{p}^{\alpha}} 
\end{equation*}
where $v_{0}$ and $v_{1}$ correspond to the minimum and maximum solar wind velocities, $f_{p}$ is the magnetic field expansion factor, and $\alpha$ is an additional numerical parameter.

The Sobol' sensitivity indices described in Eqns.~\eqref{first-order-indices}--\eqref{total-order-indices} for the WS model parameters $v_{0}, v_{1}, \alpha$ can be computed analytically (symbolically) if we assume the model parameters are independent and have uniform priors. In contrast, for the PFSS, WSA, and HUX models, the sensitivity indices can only be approximated numerically via MC integration, see Section~\ref{sec:mc-sobol}. 
We set the priors to be uniform with ranges listed in Table~\ref{tab:uncertain-parameters}. Table~\ref{tab:wang-sheeley-indicies} shows the Sobol' sensitivity indices of the three model parameters $v_{0}, v_{1}, \alpha$ for $f_{p} = 10, 10^{2}, 10^{4}$. Larger $f_{p}$ corresponds to slower solar wind velocity, in which case $v_{0}$ becomes more influential, and $v_{1}$ becomes less influential. The second-order indices show that $v_{0}$ and $v_{1}$ do not interact and that $\alpha$'s interaction with $v_{0}$ and $v_{1}$ is minor compared to the first-order indices. By the first- and total-order indices of $\alpha$, we can conclude that it is the most influential parameter independent of $f_{p}$ (in comparison to $v_{0}$ and $v_{1}$) which agrees with the ordering in the WSA model, see Section~\ref{sec:gsa-results}. 
\noindent 
\begin{table}
\caption{The analytically computed Sobol' sensitivity indices of the WS model for $f_{p} = 10, 10^{2}, 10^{4}$. The results show that $\alpha$ is the most influential parameter (in comparison to $v_{0}$ and $v_{1}$). Additionally, the indices indicate that $v_{0}$ is more influential when $f_{p}$ is high and that $v_{1}$ is more influential when $f_{p}$ is low.}
\centering
\begin{tabular}{c c c c c c c c c c} 
$f_{p}$ & $S_{v_{0}}$ & $S_{v_{1}}$ & $S_{\alpha}$ & $S_{v_{0}, v_{1}}$ & $S_{v_{0}, \alpha}$ & $S_{v_{1}, \alpha}$ & $T_{v_{0}}$ & $T_{v_{1}}$ & $T_{\alpha}$\\
\hline 
10 & 0.061 & 0.383 & 0.512 & 0 & 0.008 & 0.034 & 0.07 & 0.417 & 0.554\\
\hline
$10^{2}$& 0.131 & 0.133 & 0.679 & 0 & 0.011 & 0.045 & 0.143 & 0.178 & 0.735\\
\hline
$10^{4}$& 0.289 & 0.036 & 0.623 & 0 & 0.01 & 0.041 & 0.3 & 0.077 & 0.674\\
\end{tabular}
\label{tab:wang-sheeley-indicies}
\end{table}

\section{Affine Invariant Ensemble Sampler (AIES) and Convergence Assessment}\label{sec:aies}
In this study, we use the affine invariant ensemble sampler (AIES) developed by~\cite{goodman_2010_emcee}, which is an adaptive ensemble extension of the original Metropolis-Hastings sampler~\cite{metropolis_1953_mcmc, hastings_1970_mcmc}. Instead of evolving a single Markov chain, AIES evolves an ensemble of chains in parameter space called \emph{walkers}.
It is computationally advantageous to evolve an ensemble of chains instead of a single Markov chain since one can exploit the parallelism of the ensemble method and reach convergence significantly faster (as measured by the integrated autocorrelation time, see Eq.~\eqref{iac-defintion}).
AIES is invariant under an affine transformation of the parameter space, which is particularly appealing for problems where the parameter scales vary by several orders of magnitude, i.e. highly anisotropic target pdfs. AIES can transform anisotropic pdfs to isotropic pdfs with an affine transformation, which is much easier to sample from. Additionally, other MCMC samplers typically require tuning many sampler hyperparameters; for example, Metropolis-Hastings has $d^2$ hyperparameters where $d$ is the number of uncertain parameters (entries of the Gaussian proposal distribution covariance). Such tuning is often infeasible when the posterior evaluations are computationally demanding, as is the case in many space weather applications. AIES addresses this challenge by having only two hyperparameters in the stretch move~\cite{goodman_2010_emcee}. One hyperparameter in AIES is the number of walkers $L$, which is required to be greater than double the number of uncertain parameters $L \geq 2d + 1$, and the other hyperparameter denoted by $a$ is related to the stretch move, which we explain next.  

The AIES stretch move is described as follows. Consider an ensemble of walkers $\{\Upsilon_{1}(\ell), \ldots, \Upsilon_{L}(\ell)\}$, where $\ell = 1, \ldots, M$ is the iteration index and $L$ is the number of walkers. The proposed next step for an arbitrary walker $\Upsilon_{k}(\ell)$ is given by
\begin{equation*}
    \Upsilon_{k}(\ell+1) = \Upsilon_{j}(\ell) + S \left(\Upsilon_{k}(\ell) - \Upsilon_{j}(\ell)\right)
\end{equation*}
where $\Upsilon_{j}(\ell)$ is a complementary walker in the ensemble chosen at random (where $j \neq k$), $S$ is a random variable with density $g(s)$ that satisfies 
$g\left(\frac{1}{s}\right) = s g \left(s\right)$. An example of such a density, proposed by~\cite{goodman_2010_emcee} and implemented in the \texttt{emcee} Python package~\cite{mackey_python_2013_emcee}, is
\begin{equation}\label{density-of-scaling-parameter}
    g(s) = \left\{
    \begin{array}{ll}
          \frac{1}{\sqrt{s}} & s \in [\frac{1}{a}, a] \\
          0 & \mathrm{otherwise},\\
    \end{array}\right.
\end{equation}
where $a>1$ can be adjusted to improve the sampler's performance and is typically set to $a=2$. Thus, the proposed next step for a given walker lies on a straight line connecting the walker's current location and another random walker in the ensemble. The acceptance probability of the next proposed step is
\begin{equation*}
    \mathbb{P}\left(\Upsilon_{k}(\ell+1)|\Upsilon_{k}(\ell)\right) = \min \left(1, S^{d-1} \frac{\pi(\Upsilon_{k}(\ell+1))}{\pi(\Upsilon_{k}(\ell))}\right),
\end{equation*}
where $S$ is the random variable with density defined in Eq.~\eqref{density-of-scaling-parameter}, $d$ is the number of uncertain parameters, and $\pi$ is the target pdf. If the proposal is rejected, then $\Upsilon_{k}(\ell+1)=\Upsilon_{k}(\ell)$.

In this study, we use the Python implementation of AIES, i.e. the \texttt{emcee} package (version 3.1.4) developed by~\cite{mackey_python_2013_emcee}, with the stretch move, $a=2$, and $L=250$ walkers. We initialize the walkers by randomly sampling a Gaussian density with the mean set to the prior mean and standard deviation set to $10^{-2}$ times the prior range. 

\textit{Markov Chain Monte Carlo Burn-in.} MCMC \textit{burn-in} refers to the period when a Markov chain exhibits initial transient behavior unrepresentative of the target pdf. It is therefore recommended to disregard the first few iterations at the beginning of the Markov chain~\cite[\S 8.4]{ralph_smith_uq_2013}. Burn-in is typically an artifact of selecting a low-probability initial condition and can also be thought of as a way to find a better initial condition. The burn-in length can be chosen by detecting the iteration where the target pdf evaluations start to plateau, which can be assessed visually (or statistically) by monitoring the likelihood evaluations and the marginal paths associated with each parameter as a function of MCMC iterations. We found that after $10^{3}$ iterations, the likelihood evaluations began to plateau, meaning the Markov chains entered a region of high probability. We, therefore, disregard the first $10^{3}$ samples in each walker, which we consider as the burn-in period. 

\textit{Markov Chain Monte Carlo Convergence Assessment.} Estimating the mean of a Markov chain (or an ensemble of Markov chains) is challenging since its samples are not independent and identically distributed (i.i.d.). This is because---by definition---each sample depends on the previous sample in a Markov chain. Therefore, samples drawn close to each other tend to be correlated. 
The MC mean estimator of an ensemble of Markov chains with $L$ walkers and $M$ iterations is an unbiased estimator, i.e. 
\begin{equation*}
    \hat{\mu} = \frac{1}{M}\sum_{\ell=1}^{M} \left(\frac{1}{L} \sum_{j=1}^{L} \Upsilon_{j}(\ell) \right) \qquad \mathrm{with} \qquad  \mathbb{V}\mathrm{ar}[\hat{\mu}] = \frac{\tau}{L M}\mathbb{V}\mathrm{ar}[\Upsilon], 
\end{equation*}
where $\tau$ is the \textit{integrated autocorrelation time} (IAT)
\begin{equation}\label{iac-defintion}
    \tau = \sum_{\ell = -\infty}^{\infty} \frac{C(\ell)}{C(0)} = 1 + 2 \sum_{\ell=1}^{\infty} \frac{C(\ell)}{C(0)}
\end{equation}
and $C(\ell) = \lim_{h \to \infty} \mathbb{C}\mathrm{ov}[\Upsilon(\ell+h), \Upsilon(h)]$ is the lag-$\ell$ autocovariance function. In practice, the IAT and the autocovariance function are estimated using a finite Markov chain of length $M$, see~\cite{mackey_python_2013_emcee} for a more detailed discussion. The larger the IAT, the more samples are needed to converge to the target pdf. 
In this study, we run the chains until their length $M$ is at least $50$ times the maximum IAT (which is computed for each parameter) as suggested by~\cite{mackey_python_2013_emcee} and compute the estimated IAT using the Python  \texttt{emcee} package~\cite{mackey_python_2013_emcee}.

\section*{Open Research}
The public repository \url{https://github.com/opaliss/Parameter_Estimation_Solar_Wind} contains a collection of Jupyter notebooks in Python 3.9 containing the code and data used in this study. The GONG synoptic maps are retrieved from \url{https://gong.nso.edu/data/magmap/crmap.html} and the ACE spacecraft observations can be found at \url{https://cdaweb.gsfc.nasa.gov/index.html/}.
\section*{acknowledgments}
O.I. and B.K. were partially supported by the National Science Foundation under Award 2028125 for ``SWQU: Composable Next Generation Software Framework for Space Weather Data Assimilation and Uncertainty Quantification". E.C. was partially supported by NASA grants 80NSSC20K1580 ``Ensemble Learning for Accurate and Reliable Uncertainty Quantification" and 80NSSC20K1275 ``Global Evolution and Local Dynamics of the Kinetic Solar Wind". P.R. acknowledges support from NASA (80NSSC18K0100, NNX16AG86G, 80NSSC18K1129, 80NSSC18K0101, 80NSSC20K1285, 80NSSC18K1201, and NNN06AA01C), NOAA (NA18NWS4680081), and the U.S. Air Force (FA9550-15-C-0001).

\clearpage 
\bibliography{Issan_bibfile}
\end{document}

%% file: ListPackages.tex
\usepackage[margin = 1.2in]{geometry}
\usepackage[T1]{fontenc}
\usepackage[english]{babel}
\usepackage{natbib}
\usepackage[colorlinks=true,linkcolor=blue]{hyperref}
\PassOptionsToPackage{unicode}{hyperref}
\PassOptionsToPackage{naturalnames}{hyperref}
\hypersetup{pdfauthor=whatever}

\usepackage{mathtools}
\usepackage{amsmath}
\usepackage{amsfonts}
\usepackage{amsthm}
\usepackage{amssymb}
\usepackage{array}
\usepackage{algorithm} 
\usepackage{algorithmicx}
\usepackage{algpseudocode}
\usepackage{subcaption}
\usepackage{siunitx}

\usepackage{color}
\usepackage{url}
\usepackage{cleveref}
\usepackage{graphicx}
\usepackage{breakcites}
\usepackage{soul} 
\usepackage{enumitem}
\usepackage[title,titletoc,toc]{appendix}

%% file: mymacros.tex
{\noindent {\textbf{Proof}:} }%
{\hfill $\Box$ \\[1ex] }

\newcommand{\bit}{\begin{itemize}}
\newcommand{\eit}{\end{itemize}}
\newcommand{\ben}{\begin{enumerate}}
\newcommand{\een}{\end{enumerate}}

